\documentclass[journal]{IEEEtran}
\usepackage{cite}
\usepackage{commath}
\usepackage{amsmath,amssymb,amsfonts}
\usepackage{scalefnt}
\usepackage{algorithmic}
\usepackage{graphicx}
\usepackage{textcomp}
\usepackage{nicefrac}
\usepackage{steinmetz}
\usepackage{float}
\usepackage{makecell}
\usepackage{longtable}
\usepackage{subfigure}
\usepackage{cuted} 
\usepackage{dblfloatfix}    
\usepackage{enumitem}
\usepackage{multirow}
\usepackage{pifont}
\usepackage{mathtools}
\usepackage[dvipsnames]{xcolor}

\ifCLASSINFOpdf
\else
\fi
\begin{document}
%
\vspace{-.5cm}
\title{Improved current reference calculation for MMCs internal energy balancing control}


\author{Daniel Westerman Spier,~\IEEEmembership{Graduate Student Member,~IEEE,} Eduardo Prieto-Araujo,~\IEEEmembership{Senior Member,~IEEE,} \\Joaquim L\'opez-Mestre  and Oriol Gomis-Bellmunt,~\IEEEmembership{Fellow,~IEEE}\vspace{-0.5cm}

}

%




\IEEEtitleabstractindextext{%
\begin{abstract}
The paper addresses an improved inner current reference calculation to be employed in the control of modular multilevel converters operating during either balanced or unbalanced conditions. The suggested reference calculation is derived based on the AC and DC additive and differential voltage components applied to the upper and lower arms of the converter. In addition, the impacts caused not only by the AC network's impedances but also by the MMC's arm impedances are also considered during the derivation of the AC additive current reference expressions. Another issue discussed in this article regards that singular voltage conditions, where the positive-sequence component is equal to the negative one, may occur not only in the AC network but also internally (within the converter's applied voltages). Several different inner current reference calculation methods are compared and their applicability during the former fault conditions is analyzed. The paper presents a detailed formulation of the inner current reference calculation and applies it to different unbalanced AC grid faults where it is shown that the presented approach can be potentially used to maintain the internal energy of the converter balanced during normal and fault conditions.
\end{abstract}
\vspace{-0.2cm}
}

\maketitle

\IEEEdisplaynontitleabstractindextext

%
\IEEEpeerreviewmaketitle

\vspace{-0.6cm}
\section{Introduction}

\IEEEPARstart{W}{\lowercase{i}}th the increasing penetration of renewable energies in the power system, more Voltage Source Converters (VSC) for High Voltage Direct Current (HVDC) have been used in applications such as long transmission links and cable lines \cite{book_oriol}. Due to its improved efficiency, easier scalability to high voltage levels and inherent redundancy, the modular multilevel converter (MMC) has become the preferred converter choice for VSC-HVDC applications \cite{visionary_paper_oriol, lesnicar_paper}. Compared to classic two- and three-level converters, controlling the MMC is more complex since it has additional degrees of freedom that can be used to improve the converter performance \cite{joan_adria,daniel_iecon,daniel_tpwd,7482715}.

For a proper operation, several magnitudes of the MMC must be regulated (i.e. the AC and DC networks currents, circulating current, sub-module capacitor voltages). Regarding the internal energy of the converter, different control strategies have been proposed to regulate the total stored energy, to balance the energies between the arms and phase-legs and to maintain similar voltage levels within the sub-module capacitors \cite{b_williams,j_pou}. Under balanced grid conditions, the energy balancing is not a major issue as all the MMC's sub-modules remain close to their nominal voltage values. Nevertheless, energy deviations may occur during power changes in the network which must be compensated to maintain the proper operation of the system \cite{enric,Enric_2,6863645}. On the other hand, during unbalanced AC grid faults, the internal energy deviations are higher due to uneven power exchanges between the converter's arms and phase-legs which must be quickly compensated to avoid tripping the converter.

Relevant reference calculation approaches and control strategies have been proposed to analyze and mitigate the effects of unbalanced AC network voltages. In \cite{sul_2}, the controllers were derived targeting the horizontal (between phase-legs) energy regulation of the converter during AC single-line-to-ground fault. Still focusing on the phase-leg balancing of the MMC, \cite{wang} improves it through AC zero-sequence voltage injection while \cite{ou} analyzes the energy dynamic response for different DC circulating current scenarios. By considering not only the horizontal balancing but also the energy transfer between the upper and lower arms, further improvements can be achieved in the transient response of the converter \cite{Leon}. References \cite{Liang,bergna_3,6877696} derive distinct methods to perform the vertical energy balance of the MMC, but they only consider the DC characteristics of the circulating current. Authors in \cite{Moez_2,PRIETOARAUJO2017424,Moez_1} employ both AC and DC components of the circulating current in their control design. In \cite{Moez_2}, the MMC's AC circulating current reference calculation is performed based only on the positive sequence of the AC grid voltage, whereas \cite{PRIETOARAUJO2017424} also uses the negative sequence component. Still, the former proposals neglected the impact that the equivalent impedance of the MMC causes in the arms' applied voltages. Such influence is considered in \cite{Moez_1}, whereby an optimization algorithm using a linear matrix inequality approach is proposed to calculate the circulating current reference.

Although the previous methods are capable of controlling the MMC under unbalanced AC network conditions, during unbalanced scenarios whereby the positive and negative components of either the AC grid voltages or the MMC's internal voltages are equal, they might result in singularities in the current reference calculation. This problem was addressed for two-level VSCs by \cite{adria}, and possible solutions were proposed for MMC applications during the former AC grid voltage sag in \cite{edu_tran}.

To the best of the authors knowledge, a control strategy that is capable of dealing with different unbalanced AC voltage sags (in particular internal singular voltage sags and singular AC network faults), while maintaining the upper and lower arms energies balanced has not been proposed yet. Another research gap considered in this paper regards the addition of the internal impedance effects on the AC additive voltage during the derivation of the MMC's circulating current references. Next, the main contributions of this paper are highlighted:

\begin{itemize}
    \item Comparison among different AC additive current reference calculations and their potential usage during singular voltage condition.
    \item The DC differential zero-sequence voltage component $U_{diff}^{0DC}$ is employed to enhance the power balancing between the upper and lower arms throughout the operation of the converter (balanced or unbalanced).
    \item Improvements in the solutions proposed in \cite{edu_tran} with the addition of $U_{diff}^{0DC}$ and the MMC's equivalent impedance.
    \item The degrees of freedom of the MMC are fully exploited by the current reference calculations.
    \item Comprehensive additive current reference calculation able to operate in any grid voltage condition.
\end{itemize}

The proposed reference calculation is compared with different methods by means of time-domain simulation results for distinct AC grid and internal singular voltage sag conditions. In addition, the effects of $U_{diff}^{0DC}$ are also analyzed according to the reference calculation approach employed.

\vspace{-0.3cm}
\section{MMC system description and modelling}
\begin{figure}[!h]
\vspace{-0.3cm}
\centerline{\includegraphics[width=0.85\linewidth]{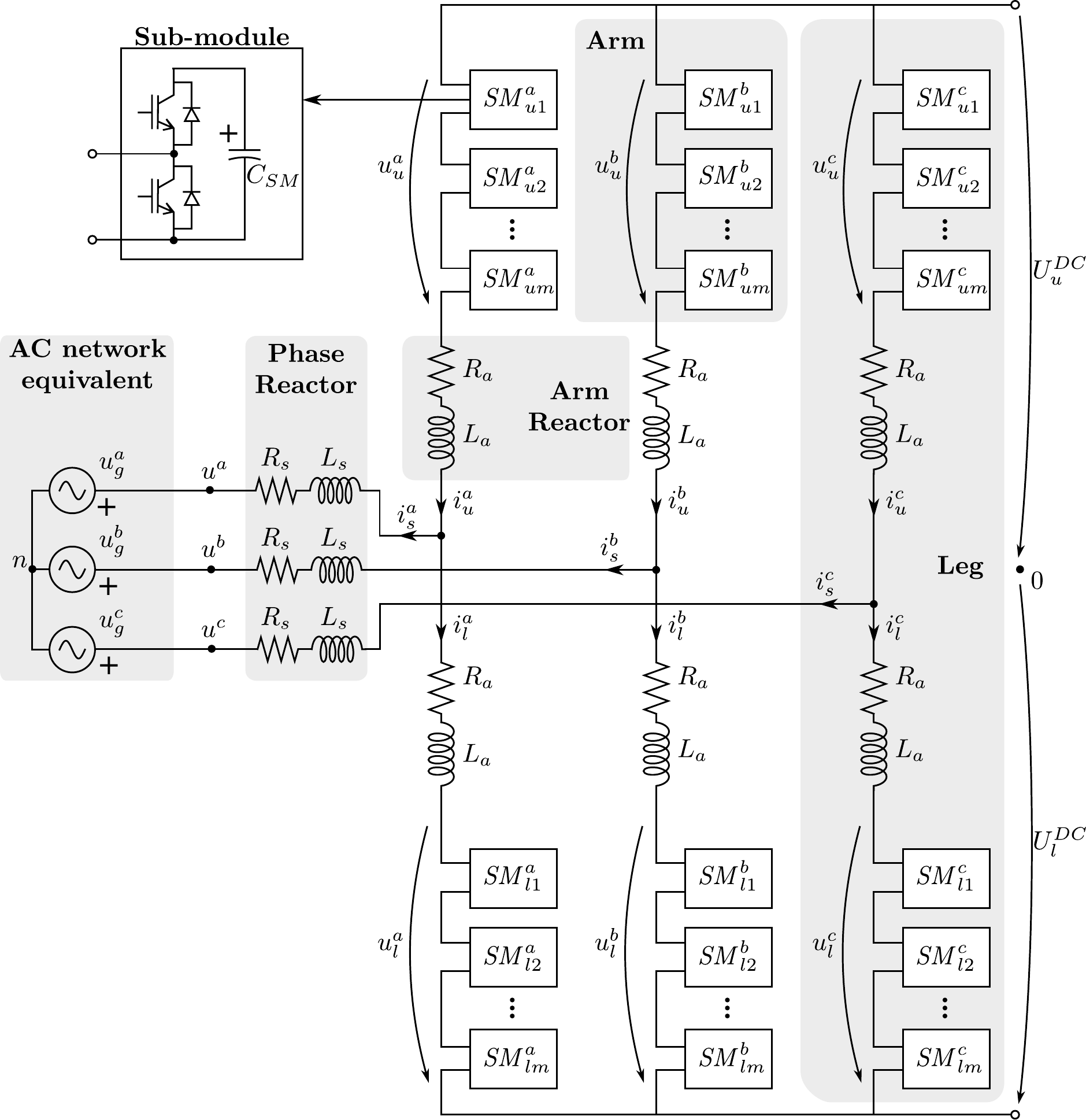}}
\vspace{-0.1cm}
\caption{Complete model of the MMC converter.}
\label{fig:MMC}
\vspace{-0.3cm}
\end{figure}

\vspace{-0.1cm}
The model of the three-phase MMC is shown in Fig. \ref{fig:MMC}. The converter consists of three legs, one per phase, where each leg has two stacks of $N_{arm}$ sub-modules (SMs), known as the upper and lower arms. The topologies employed in the sub-modules varies according to the application requirements, where the half-bridge structure is widely used due to its simplicity and lower costs \cite{mmc_book}.

For steady-state modelling purposes, the phasor notation $\underline{X}^k = X_r^k + j X_i^k = X^k \phase{\theta^k}$ will be adopted, with $x(t) = X^k \text{Re}\{e^{j( \omega t + \theta^k)}\}$ and $k \in \{a,b,c\}$. Therefore, the main quantities for each phase are: the AC grid voltages $\underline{U}_{g}^k$; the upper and lower arms applied voltages $\underline{U}_{u,l}^k$; the upper and lower DC grid voltages $U_{u,l}^{DC}$; the voltage between the 0 DC reference node and the neutral $n$ of the AC three-phase system $\underline{U}_{0n}$; the upper and lower arm currents $\underline{I}_{u,l}^k$; the AC grid current $\underline{I}_{s}^k$; the arm impedances $R_a$ and $L_a$; the grid equivalent resistance and inductance $R_s$ and $L_s$; and the sub-module capacitors $C_{SM}$. To ease the understanding of the MMC circuit and the derivation of the control strategies, the following quantities are defined:
\vspace{-0.1cm}
\begin{equation}
\small
\left\{  \begin{array}{l}
\underline{U}_{diff} \triangleq \dfrac{1}{2}(-\underline{U}_{u}+\underline{U}_{l})\\
\underline{U}_{sum} \triangleq \underline{U}_{u}+\underline{U}_{l}\\
\underline{I}_{sum} \triangleq \dfrac{1}{2}(\underline{I}_{u}+\underline{I}_{l})\\
\end{array}\right. \quad 
\left\{  \begin{array}{l}
Z_{arm}\triangleq R_a+j\omega L_a\\
Z_{s}\triangleq R_s+j\omega L_s \\
Z_{eq}\triangleq Z_s + \dfrac{Z_a}{2} \\
\end{array}\right.
\label{eq:varchange2}
\end{equation}

\noindent with $k \in \{a,b,c\}$ and $\underline{U}_{diff}^k$, $\underline{U}_{sum}^k$ and $\underline{I}_{sum}^k$ as the differential and additive applied voltages and circulating currents of the converter, respectively. As it will be described later, the additive currents perform an important role to regulate the power transfer from/to the HVDC network to/from the MMC and to maintain the internal power of the converter balanced, which can be achieved by exchanging power among phase-legs and between the upper and lower arms. 

\vspace{-0.5cm}
\section{MMC control system}
\begin{figure}[!h]
\vspace{-0.3cm}
\centerline{\includegraphics[width=1\linewidth]{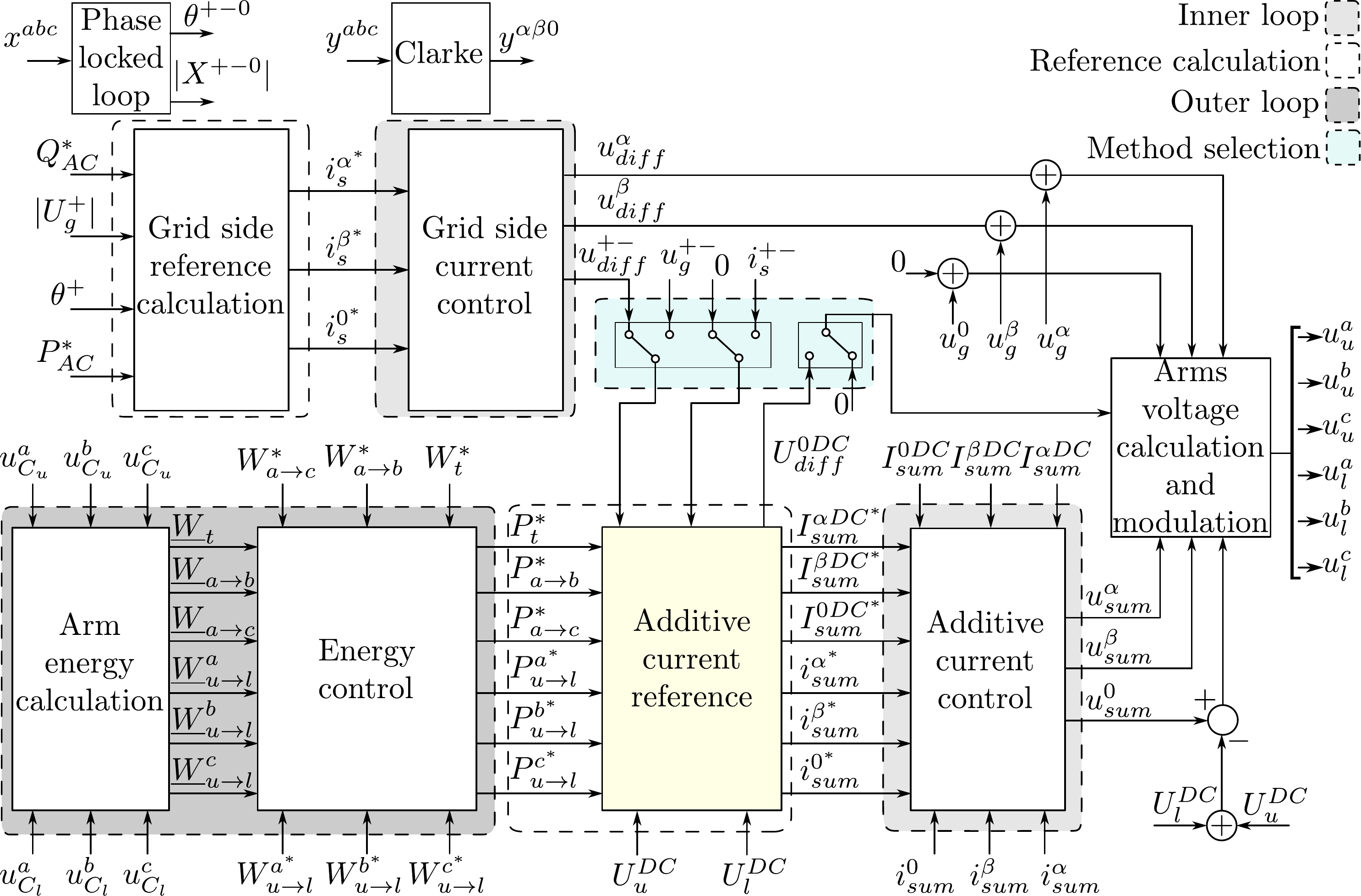}}
\vspace{-0.1cm}
\caption{Overall control scheme of the MMC converter for grid-following applications.}
\label{fig:control_scheme}
\vspace{-0.5cm}
\end{figure}

\vspace{-0.1cm}
In this section, the overall control system to regulate the MMC and the methodology to calculate the additive and AC network current references are presented. The employed control scheme, shown in Fig. \ref{fig:control_scheme}, uses the design procedures derived in \cite{PRIETOARAUJO2017424} and can be divided into two main parts: the AC grid current control and the circulating current control. For the AC grid current stage, the current references are calculated based on the active and reactive power set-points required by the Transmission System Operator (TSO) and the magnitude of the positive-sequence component of the AC network voltage. Then, such references are employed into the grid side current control loops. The energy controllers are designed in order to maintain the internal energy balance of the converter. This is achieved through six different control loops which regulates the MMC's total internal energy $E_t$, the energy difference between the converter's phase-legs $E_{a\to b}$ and $E_{a\to c}$ and the energy mismatch between the converter's arms $E_{u\to l}^k$. Such energy regulators result in the power references necessary to calculate the AC and DC inner current set-points, highlighted in yellow in Fig. \ref{fig:control_scheme}. These current references can be obtained through different methods according to the quantities that are used in the calculation (see method selection in Fig. \ref{fig:control_scheme} and Sections \ref{sec:comparison} and \ref{sec:proposed}), which are later tracked by the additive current controllers.

For the sake of completeness, the reference calculation procedures for the AC grid currents and for the DC component of the additive currents are briefly described next, which more details can be found in \cite{PRIETOARAUJO2017424}. On the other hand, a comprehensive analysis of different methods to calculate the current references for the AC component of the additive current is given. In addition, such approaches are compared regarding their applicability during faults and exploitation of the MMC's degrees of freedom.

\vspace{-0.5cm}

\subsection{AC network current reference calculation}
Under balanced conditions, the AC grid currents present a symmetrical profile. However, during unbalanced AC voltage sags, the three-phase system may have different voltage levels for each phase (due to the presence of negative-sequence components), which might result in unbalanced currents circulating through the AC network. In general, the AC grid current references are calculated by considering only the positive-sequence of the AC grid voltages, either for balanced or unbalanced scenarios \cite{Akagi2007InstantaneousPT}. 

\vspace{-0.5cm}
\subsection{Additive current reference calculation}
Generally, the DC components of the additive current are used to regulate the power transfer among the phase-legs of the converter. Whereas the AC components are employed to control the power exchanged between the MMC's upper and lower arms (vertical balancing) \cite{Leon}.

\subsubsection{DC component of the additive current}
The DC terms of the additive current can be applied in the regulation of the energy exchanged horizontally. In addition, notch filters must be used in order to eliminate the line and double-line frequency power components coming from the energy controllers  \cite{PRIETOARAUJO2017424}.

\subsubsection{AC component of the additive current}
To calculate the AC additive current references it is necessary to obtain a mathematical expression relating the power difference between the MMC's upper and lower halves with their respective applied voltages. At one hand, the arm's applied voltages can be considered to be equal to the AC network voltage (assuming that the equivalent impedance of the converter is small). The main advantages of such approach regards its simplicity and straightforwardness, since it uses the measurements from the AC system. However, this method presents a discontinuity when the positive- and negative-sequence components of the AC grid voltages are equal or almost equal (singular condition) \cite{edu_tran}. As a consequence, the additive current references saturate, compromising the vertical energy balancing controller. Another candidate solution would to employ the differential voltages resultant from the grid side current control as the arm's applied voltages, but as it will be demonstrated later, this strategy also fails if the internal voltages present singular characteristics.

\vspace{-0.2cm}
\section{Comparison among different AC additive current reference calculation strategies} \label{sec:comparison}

In this section, different methods to calculate the AC additive current references are presented. These approaches vary according to the voltages that are considered to be applied into the MMC's arms (AC grid or differential voltages). For balanced and several unbalanced network conditions, the distinct strategies can maintain the converter stable and provide proper current references. However, certain unbalanced AC grid voltage sags may lead to internal singular voltage conditions that must also be addressed during the derivation of the additive current references to avoid discontinuities of the system. 

\vspace{-0.3cm}
\subsection{Internal singular voltage sag analysis} \label{sec:internal}
An internal singular voltage event is defined when the positive-sequence component of the AC differential voltage is equal to the negative one $(\underline{U}_{diff}^+ = \underline{U}_{diff}^-)$. In order to derive the expression for such fault, let's first assume that the AC differential voltages are calculated as \cite{Enric_2}

\begin{subequations}
\vspace{-0.2cm}
\small
\begin{equation}
\underline{U}_{diff}^{+} =   \underline{U}_{g}^{+} + \underline{Z}_{eq} \underline{I}_s^{+} 
\label{eq:KVL_pos}
\end{equation}
\begin{equation}
\underline{U}_{diff}^{-} =   \underline{U}_{g}^{-} + \underline{Z}_{eq} \underline{I}_s^{-} 
\label{eq:KVL_neg}
\end{equation}
\label{eq:KVL}
\vspace{-0.5cm}
\end{subequations}

The scenario where $\underline{U}_{diff}^+ = \underline{U}_{diff}^-$ is obtained by equating \eqref{eq:KVL_pos} and \eqref{eq:KVL_neg}. As a result, an expression that describes in which condition the converter's applied voltages present singular behavior is given as follows

\vspace{-0.2cm}
\begin{equation}
\small
\underline{U}_{g}^{-} = \underline{U}_{g}^{+} + \underbrace{\underline{Z}_{eq}\left(\underline{I}_s^{+} -\underline{I}_s^{-} \right) }_\text{internal factor} 
\label{eq:int_sing}
\end{equation}

As mentioned in Section III-A, the AC grid controllers are designed to inject only positive-sequence current component into the grid, thus $\underline{I}_s^- = 0$. Furthermore, it can be observed that this type of fault not only depends on the AC grid voltage characteristics but is also affected by the interaction between the MMC's equivalent impedance and the AC grid currents. In this paper, it is considered that the internal factor is constant, since the controllers will keep injecting the same positive-sequence current levels into the AC grid throughout the converter's operation.

\vspace{-0.3cm}
\subsection{Method 0 - Initial approach $\underline{U}_{u,l}^k = \underline{U}_{g}^k$}

This methodology is the most straightforward one to calculate the AC additive current references of the MMC. It assumes that the equivalent impedance of the converter is small and by doing so, the AC arm voltage can be considered equal to the AC grid voltage. However, during AC grid faults where the positive-sequence voltage component is equal to the negative one this method fails \cite{PRIETOARAUJO2017424}. Such condition arises because the current reference calculation will present a discontinuity in this operating point. As a consequence, it will try to impose very high AC additive currents to circulate through the MMC, which must be disconnected to avoid damaging the converter.

\vspace{-0.3cm}

\subsection{Method 1 - Method 0 considering $U_{diff}^{0DC}$}
This reference calculation applies the techniques developed in \cite{edu_tran} employing $U_{diff}^{0DC}$. Next, the method's working principle along with the calculation and regulation of $U_{diff}^{0DC}$ are described.

\subsubsection{Working principle}
This method removes one of the degrees of freedom from the AC additive current that does not contribute to the power exchanged within the MMC arms. By doing so, the discontinuity is avoided but there will be a sustained constant energy deviation between the phase-legs of the converter throughout the fault. This approach can be further improved by replacing the degree of freedom that was lost with the DC differential zero-sequence voltage component $U_{diff}^{0DC}$.

\subsubsection{Regulation of $U_{diff}^{0DC}$}
During balanced conditions, this voltage is equal to zero as both  upper and lower arms have the same DC voltage level. Under transients, on the other hand, $U_{diff}^{0DC}$ is different than zero and its magnitude can be applied into the MMC's arms to improve the energy balancing between them. Such degree of freedom can be obtained as

\begin{figure}[!h]
\vspace{-0.3cm}
\centerline{\includegraphics[width=0.7\linewidth]{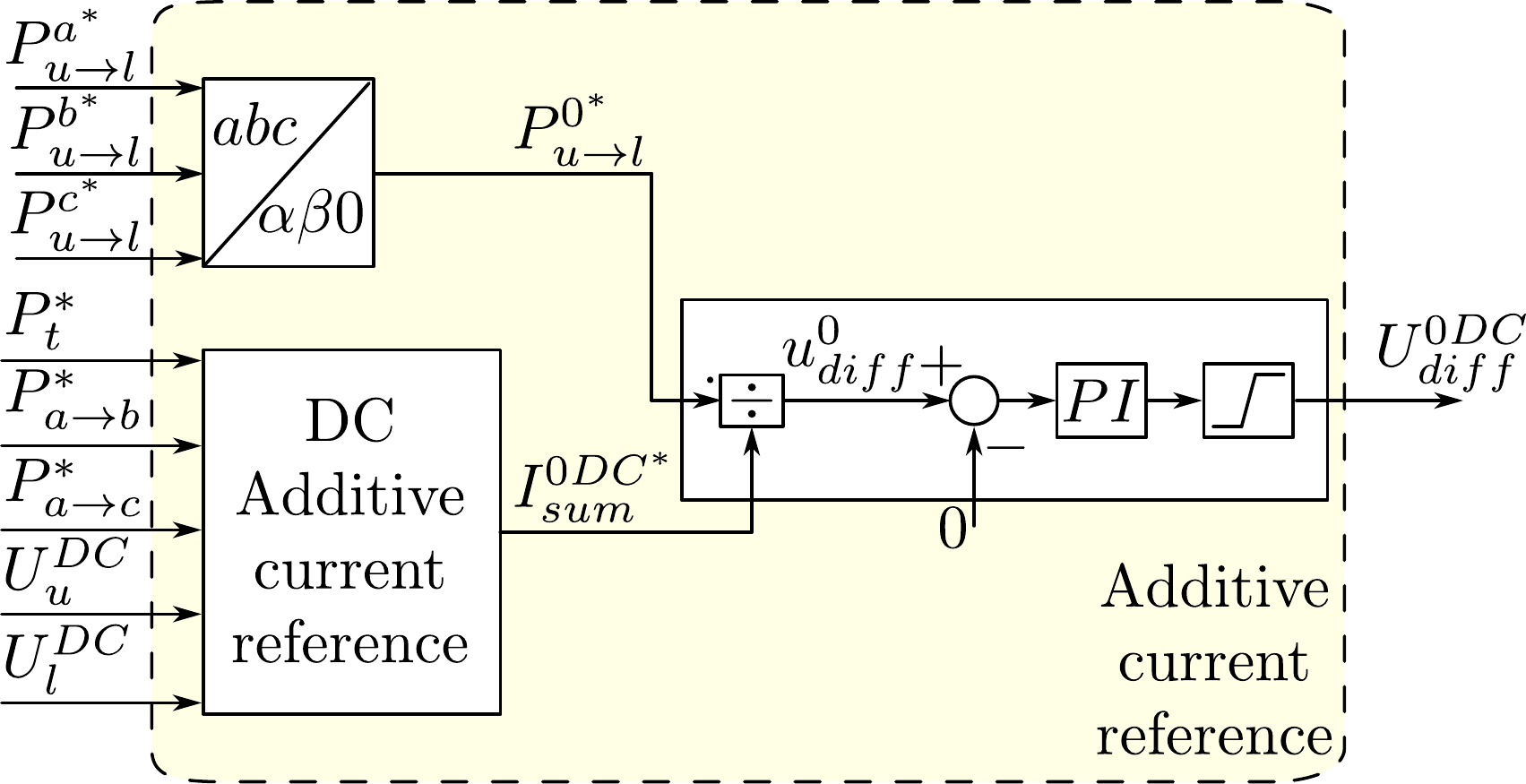}}
\vspace{-0.1cm}
\caption{$U_{diff}^{0DC}$ control structure.}
\label{fig:Udiff_0DC_control}
\vspace{-0.3cm}
\end{figure}

\vspace{-0.1cm}
\begin{equation}
\small
    U_{diff}^{0DC} = \dfrac{P_{u \to l}^a +P_{u \to l}^b + P_{u \to l}^c}{3I_{sum}^{0DC}}
    \label{eq:Udiff_0DC}
    \vspace{-0.1cm}
\end{equation}

\noindent where $P_{u \to l }^{abc}$ are the power difference between the upper and lower arms and $I_{sum}^{0DC} \neq 0$ (to avoid discontinuities). By using this degree of freedom, the energy deviations are eliminated and controlled power can be transmitted within the MMC's arms, enhancing the converter response during singular AC grid voltage conditions (see Section VI). In Fig. \ref{fig:Udiff_0DC_control}, the control structure employed in the regulation of $U_{diff}^{0DC}$ is depicted. As aforementioned, the unregulated value of $U_{diff}^{0DC}$ is calculated based on the equation shown above, and compared to its desired magnitude (set to be equal to zero), resulting in the error \textbf{\textit{e}}. This error goes to a PI controller, which is designed in order to quickly compensate any voltage disparity caused by the zero-sequence power mismatch between the upper and lower arms of the converter. By doing so, the sustained energy deviations observed when other control methods are used, as it was pointed out by \cite{edu_tran}, can be eliminated. The controller gains employed in this paper are set to be equal to $k_p=0.25$ and $k_i=12$, whereby the gain selection was done based on the response of $U_{diff}^{0DC}$ for different AC and internal singular voltage sag conditions. Finally, a saturation block is added as a safety factor in order to prevent high values of $U_{diff}^{0DC}$ which would result in overmodulations. 

\vspace{-0.3cm}
\subsection{Method 2 - Arm voltages equal to the DC and AC differential voltages $\underline{U}_{u,l}^{+-} = \underline{U}_{diff}^{+-}$ + $U_{diff}^{0DC}$}

An alternative solution to the former problems would be to replace the AC grid voltages by the positive- and negative-sequence components of the internal differential voltages of the converter. Consequently, the MMC equivalent impedance will not be neglected, having a more realistic AC voltage level in the converter's arms. For this strategy, however, the differential voltage controllers should be fast enough in order to avoid inaccuracies due to the interactions between the two regulators \cite{harnefors}. Nevertheless, even if such requirements are fulfilled, in a situation where the MMC's internal differential voltages are equal, $U_{diff}^{+} = U_{diff}^-$, the discontinuity will also occur and the vertical energy balancing of the converter will be compromised.  
\vspace{-0.3cm}
\subsection{Method 3 - Method 1 with $\underline{U}_{u,l}^{+-} = \underline{U}_{diff}^{+-} + U_{diff}^{0DC}$}

Now, the principles in Method 1 are extended by considering that the arm's applied voltages are equal to the differential ones. If only the AC differential terms are employed, during fault event where $U_{diff}^{+} = U_{diff}^-$, the continued energy deviation would also be observed and would be compensated with the zero sequence DC differential voltage. However, as it will be shown in Section \ref{sec:Results}, the energy drifts will cause the $U_{diff}^{0DC}$ controllers to saturate due to the limited voltage application range. Consequently, the vertical energy regulators will not be able to compensate the power transferred within the converter's arms, leading to the eventual disconnection of the system.

\vspace{-0.3cm}

\section{Method 4 - Proposed approach considering the additive and differential voltage components in the arm} \label{sec:proposed}

The previous methods present distinct strategies to calculate the AC additive current by using different quantities as the applied voltages in the MMCs arms. But, they share the same characteristic where the additive voltage effects in the arm are neglected. At one hand, it seems that such voltage cannot be part of the AC additive current references without an iterative control method. However, using a simple mathematical substitution, which does not require any optimization, iterative control or violate any constraint imposed in the steady-state analysis, it is possible to obtain an expression that can be employed as such reference calculation for any AC grid or arm singular voltage sag condition. To do so, let's first consider that the applied voltages in the six arms are

\vspace{-0.2cm}
\begin{subequations}
\small
\begin{equation}
\underline{U}_{u,l}^{k} = \mp \underline{U}_{diff}^{k} +\dfrac{\underline{U}_{sum}^{k}}{2}
\label{eq:U_upper_simp}
\end{equation}
\vspace{-0.5cm}
\begin{align}
\label{eq:U_ul_abc}
&u_{u,l}^{k}(t) =\sqrt{2}\Bigg(
\mp  U_{diff}^{+}\cos\left(\omega t +\theta_{diff}^{+} + \alpha^k\right) \mp  \\& \nonumber \mp  U_{diff}^{-}\cos\left(\omega t +\theta_{diff}^{-} - \alpha^k\right)  +\dfrac{U_{sum}^{+}}{2}\cos{\left(\omega t + \theta_{sum}^{+} + \alpha^k \right)} + \\& \nonumber+ \dfrac{U_{sum}^{-}}{2}\cos{\left(\omega t + \theta_{sum}^{-} -\alpha^k \right)}\Bigg)  +\left(\dfrac{U_{sum}^{kDC}}{2}\mp U_{diff}^{0DC}\right)
\end{align}
\end{subequations}

\noindent where the sign $\mp$ indicates that the differential voltage components for the upper arms are negative, $u_u^k=-u_{diff}^k + \frac{u_{sum}^k}{2}$, while for the lower arms the differential voltages are positive, $u_l^k=u_{diff}^k + \frac{u_{sum}^k}{2}$. In addition, $\alpha^a = 0, \alpha^b = -\frac{2\pi}{3}, \alpha^c =\frac{2\pi}{3}$ and  $k\in \{a,b,c\}$. $\theta_{diff}^{+}$ and $\theta_{diff}^{-}$ are the phase-angles of the positive and negative sequence components of the differential voltages, whereas $\theta_{sum}^{+}$ and $\theta_{sum}^{-}$ are the phase-angles for the positive- and negative-sequence additive voltages. The upper and lower arms currents can be described as follows

\vspace{-0.1cm}
\begin{subequations}
\small
\begin{equation}
\underline{I}_{u,l}^{k} = \pm \dfrac{\underline{I}_{s}^{k}}{2} +\underline{I}_{sum}^{k}
\label{eq:I_ul_s}
\end{equation}
\vspace{-0.5cm}
\begin{align}
&i_{u,l}^{k}(t) =\sqrt{2}\Bigg(
\pm \dfrac{I_{s}^{+}}{2}\cos\left(\omega t +\phi_{s}^{+} + \alpha^k \right) \pm  \nonumber\\& \pm  \dfrac{I_{s}^{-}}{2}\cos\left(\omega t + \phi_{s}^{-} -\alpha^k \right) +  I_{sum}^{+}\cos{\left(\omega t + \phi_{sum}^{+} + \alpha^k \right)}+\nonumber\\& + I_{sum}^{-}\cos{\left(\omega t + \phi_{sum}^{-} -\alpha^k \right)}\Bigg)+ I_{sum}^{kDC}
\label{eq:Iul_abc}
\end{align}
\end{subequations}
\begin{equation}
\small
P_{u\to l}^{k}= \underline{U}_{u}^{k}\underline{I}_{u}^{k}-\underline{U}_{l}^{k}\underline{I}_{l}^{k},  \quad k\in \{a,b,c\}
\label{eq:Pul_tot_abc}
\end{equation}

\begin{figure*}[!t]
\footnotesize
\begin{subequations}
\begin{equation}
P_{u\to l}^{a}= \begin{bmatrix}
U_{diff}^{+} \\
U_{diff}^{-} \\
I_{s}^{+} \\
I_{s}^{-} \\
\end{bmatrix}^T
\begin{bmatrix}
-2\cos{(\theta_{diff}^{+} - \phi_{sum}^{+}}) & -2\cos{(\theta_{diff}^{+}  - \phi_{sum}^{-}})  & 0 & 0 \\
-2\cos{(\theta_{diff}^{-} - \phi_{sum}^{+}}) & -2\cos{(\theta_{diff}^{-} - \phi_{sum}^{-}})  & 0 & 0 \\
0 & 0 & \dfrac{\cos{(\theta_{sum}^{+} - \phi_{s}^{+})}}{2} & \dfrac{\cos{(\theta_{sum}^{-} - \phi_{s}^{+})}}{2} \\
0 & 0 & \dfrac{\cos{(\theta_{sum}^{+} - \phi_{s}^{-})}}{2}  &  \dfrac{\cos{(\theta_{sum}^{-} - \phi_{s}^{-})}}{2}\\
\end{bmatrix}
\begin{bmatrix}
I_{sum}^{+}\\ 
I_{sum}^{-}\\
U_{sum}^{+}\\ 
U_{sum}^{-}
\end{bmatrix}  -2U_{diff}^{0DC}I_{sum}^{aDC} 
\label{eq:Pu_l_a}
\end{equation}

\begin{align}
P_{u\to l}^{b}= \begin{bmatrix}
U_{diff}^{+} \\
U_{diff}^{-} \\
I_{s}^{+} \\
I_{s}^{-} \\
\end{bmatrix}^T
\begin{bmatrix}
-2\cos{(\theta_{diff}^{+} - \phi_{sum}^{+})} & -2\cos{(\theta_{diff}^{+}  - \phi_{sum}^{-} + \frac{2\pi}{3})}  & 0 & 0 \\
-2\cos{(\theta_{diff}^{-} - \phi_{sum}^{+} - \frac{2\pi}{3} )} & -2\cos{(\theta_{diff}^{-} - \phi_{sum}^{-})}  & 0 & 0 \\
0 & 0 & \dfrac{\cos{(\theta_{sum}^{+} - \phi_{s}^{+})}}{2} & \dfrac{\cos{(\theta_{sum}^{-} - \phi_{s}^{+}-\frac{2\pi}{3})}}{2} \\
0 & 0 & \dfrac{\cos{(\theta_{sum}^{+} - \phi_{s}^{-} + \frac{2\pi}{3})}}{2}  &  \dfrac{\cos{(\theta_{sum}^{-} - \phi_{s}^{-})}}{2}\\
\end{bmatrix}
\begin{bmatrix}
I_{sum}^{+}\\ 
I_{sum}^{-}\\
U_{sum}^{+}\\ 
U_{sum}^{-}
\end{bmatrix}- \\ \nonumber -2U_{diff}^{0DC}I_{sum}^{bDC} 
\label{eq:Pu_l_b}
\end{align}

\begin{align}
P_{u\to l}^{c}= \begin{bmatrix}
U_{diff}^{+} \\
U_{diff}^{-} \\
I_{s}^{+} \\
I_{s}^{-} \\
\end{bmatrix}^T
\begin{bmatrix}
-2\cos{(\theta_{diff}^{+} - \phi_{sum}^{+})} & -2\cos{(\theta_{diff}^{+}  - \phi_{sum}^{-} - \frac{2\pi}{3})}  & 0 & 0 \\
-2\cos{(\theta_{diff}^{-} - \phi_{sum}^{+} + \frac{2\pi}{3} )} & -2\cos{(\theta_{diff}^{-} - \phi_{sum}^{-})}  & 0 & 0 \\
0 & 0 & \dfrac{\cos{(\theta_{sum}^{+} - \phi_{s}^{+})}}{2} & \dfrac{\cos{(\theta_{sum}^{-} - \phi_{s}^{+}+\frac{2\pi}{3})}}{2} \\
0 & 0 & \dfrac{\cos{(\theta_{sum}^{+} - \phi_{s}^{-} - \frac{2\pi}{3})}}{2}  &  \dfrac{\cos{(\theta_{sum}^{-} - \phi_{s}^{-})}}{2}\\
\end{bmatrix}
\begin{bmatrix}
I_{sum}^{+}\\ 
I_{sum}^{-}\\
U_{sum}^{+}\\ 
U_{sum}^{-}
\end{bmatrix}- \\ \nonumber -2U_{diff}^{0DC}I_{sum}^{cDC} 
\label{eq:Pu_l_c}
\end{align}

\label{eq:Pu_l_abc}
\end{subequations}
\vspace*{-0.35cm}
\hrulefill
\vspace*{-0.5cm}
\end{figure*}
\noindent where similarly to the arms' applied voltages, the sign $\pm$ is an indication that $i_u^k = \frac{i_{s}^k}{2} + i_{sum}^k$ and $i_l^k = -\frac{i_{s}^k}{2} + i_{sum}^k$. Furthermore, $\phi_{s}^{+}$ and $\phi_{sum}^{+}$ are the phase-angles of the positive-sequence additive and AC grid currents, respectively, while $\phi_{s}^{-}$ and $\phi_{sum}^{-}$ are the phase-angles of the negative-sequence current components. Having defined the upper and lower arms voltages and currents with the additive and differential quantities, it is possible to describe the power difference between the upper and lower arms as

Replacing \eqref{eq:U_ul_abc} and \eqref{eq:Iul_abc} in \eqref{eq:Pul_tot_abc}, the power differences are obtained in matrix form as shown in \eqref{eq:Pu_l_abc}. It can be observed that the left matrix consists of the AC grid currents and differential voltages whereas the far right matrix contains the additive voltages and currents. Furthermore, the phase-angle differences for each power element indicates an interaction between the additive and differential quantities. Although \eqref{eq:Pul_tot_abc} can express the power transfer between the upper and lower arms of the converter, employing it in a control strategy to calculate the AC components additive current references might be challenging as it would require iterative calculation methods. 

A simple approach would be to neglect the additive voltages and to consider only the differential terms in the reference calculation. However, as discussed in Section IV, this approximation fails during internal singular voltage sag conditions. In order to overcome such issue and to increase the operating range of the converter, the proposed reference calculation substitute $\underline{U}_{sum}^{+-}$ by an equation containing the arm impedance and the additive current, yielding

\vspace{-0.1cm}
\begin{subequations}
\small
\begin{equation}
    \underline{U}_{sum}^{+} =-2 Z_{arm}I_{sum}^{+}\phase{\rho+\phi_{sum}^{+}}
\end{equation}
\begin{equation}
\underline{U}_{sum}^{-} =-2 Z_{arm}I_{sum}^{-}\phase{\rho+\phi_{sum}^{-}}
\end{equation}
\label{eq:Usum_new}
\vspace{-0.3cm}
\end{subequations}

\noindent where $\rho$ is the phase-angle of the arm impedance $\underline{Z}_{arm}$. Replacing the additive voltages in \eqref{eq:Pu_l_abc} with the new expressions from \eqref{eq:Usum_new}, the final equations for the vertical power transfer are obtained and expressed in \eqref{eq:Pu_l_abc_simp}.

\begin{figure*}[!t]
\footnotesize
\begin{subequations}
\begin{equation}
P_{u\to l}^{a}= \begin{bmatrix}
U_{diff}^{+} \\
U_{diff}^{-} \\
I_{s}^{+} \\
I_{s}^{-} \\
\end{bmatrix}^T
\begin{bmatrix}
-2\cos{(\theta_{diff}^{+} - \phi_{sum}^{+})} & -2\cos{(\theta_{diff}^{+}  - \phi_{sum}^{-})}\\
-2\cos{(\theta_{diff}^{-} - \phi_{sum}^{+})} & -2\cos{(\theta_{diff}^{-} - \phi_{sum}^{-})} \\
-Z_{arm}\cos(\rho + \phi_{sum}^{+}-\phi_{s}^{+}) &  -Z_{arm}\cos(\rho + \phi_{sum}^{-} - \phi_{s}^{+})\\
-Z_{arm}\cos(\rho  + \phi_{sum}^{+}-\phi_{s}^{-}) & -Z_{arm}\cos(\rho + \phi_{sum}^{-}-\phi_{s}^{-})\\
\end{bmatrix}
\begin{bmatrix}
I_{sum}^{+}\\ 
I_{sum}^{-}\\
\end{bmatrix}  -2U_{diff}^{0DC}I_{sum}^{aDC} 
\label{eq:Pu_l_a_simp}
\end{equation}

\begin{equation}
P_{u\to l}^{b}= \begin{bmatrix}
U_{diff}^{+} \\
U_{diff}^{-} \\
I_{s}^{+} \\
I_{s}^{-} \\
\end{bmatrix}^T
\begin{bmatrix}
-2\cos{(\theta_{diff}^{+} - \phi_{sum}^{+})} & -2\cos{(\theta_{diff}^{+}  - \phi_{sum}^{-}+ \frac{2\pi}{3})}\\
-2\cos{(\theta_{diff}^{-} - \phi_{sum}^{+}-\frac{2\pi}{3})} & -2\cos{(\theta_{diff}^{-} - \phi_{sum}^{-})} \\
-Z_{arm}\cos(\rho + \phi_{sum}^{+}-\phi_{s}^{+}) &  -Z_{arm}\cos(\rho + \phi_{sum}^{-} - \phi_{s}^{+}+\frac{2\pi}{3})\\
-Z_{arm}\cos(\rho  + \phi_{sum}^{+}-\phi_{s}^{-}-\frac{2\pi}{3}) & -Z_{arm}\cos(\rho + \phi_{sum}^{-}-\phi_{s}^{-})\\
\end{bmatrix}
\begin{bmatrix}
I_{sum}^{+}\\ 
I_{sum}^{-}\\
\end{bmatrix}  -2U_{diff}^{0DC}I_{sum}^{bDC} 
\label{eq:Pu_l_b_simp}
\end{equation}

\begin{equation}
P_{u\to l}^{c}= \begin{bmatrix}
U_{diff}^{+} \\
U_{diff}^{-} \\
I_{s}^{+} \\
I_{s}^{-} \\
\end{bmatrix}^T
\begin{bmatrix}
-2\cos{(\theta_{diff}^{+} - \phi_{sum}^{+})} & -2\cos{(\theta_{diff}^{+}  - \phi_{sum}^{-}- \frac{2\pi}{3})}\\
-2\cos{(\theta_{diff}^{-} - \phi_{sum}^{+}+\frac{2\pi}{3})} & -2\cos{(\theta_{diff}^{-} - \phi_{sum}^{-})} \\
-Z_{arm}\cos(\rho + \phi_{sum}^{+}-\phi_{s}^{+}) &  -Z_{arm}\cos(\rho + \phi_{sum}^{-} - \phi_{s}^{+}-\frac{2\pi}{3})\\
-Z_{arm}\cos(\rho  + \phi_{sum}^{+}-\phi_{s}^{-}+\frac{2\pi}{3}) & -Z_{arm}\cos(\rho + \phi_{sum}^{-}-\phi_{s}^{-})\\
\end{bmatrix}
\begin{bmatrix}
I_{sum}^{+}\\ 
I_{sum}^{-}\\
\end{bmatrix}  -2U_{diff}^{0DC}I_{sum}^{cDC} 
\label{eq:Pu_l_c_simp}
\end{equation}

\label{eq:Pu_l_abc_simp}
\end{subequations}
\vspace*{-0.15cm}
\hrulefill
\vspace*{-0.5cm}
\end{figure*}

Comparing \eqref{eq:Pu_l_abc} and \eqref{eq:Pu_l_abc_simp}, it can be noted that the substitution does not change the number of terms in the new power transfer equations; thus, the degrees of freedom of the converter are still being fully exploited. The three vertical power quantities $\left(P_{u\to l}^{a}, P_{u\to l}^{b}, P_{u\to l}^{c}\right)$ are adjusted based on four parameters  $\left(I_{sum}^{+},I_{sum}^{-}, \phi_{sum}^{+}, \phi_{sum}^{-} \right)$, since the AC grid current and the differential voltage magnitudes are given values that are regulated independently of the internal powers. By choosing that the reactive component of the positive additive current is equal to zero $(\sin(\phi_{sum}^+) = 0)$, \eqref{eq:Pu_l_abc_simp} can be reduced to 

\vspace{-0.4cm}
\begin{equation}
\small 
\hspace{-0.2cm}
\underbrace{\begin{bmatrix}
P_{u\to l}^{a}\\P_{u\to l}^{b}\\P_{u\to l}^{ac}
\end{bmatrix}}_\text{P}
 = 
\underbrace{\begin{bmatrix}
M_{11}&M_{12}&M_{13}\\
M_{21}&M_{22}&M_{23}\\
M_{31}&M_{32}&M_{33}\\
\end{bmatrix}}_\text{M}
\underbrace{\begin{bmatrix}
I_{sum}^{-}\cos{\phi_{sum}^{-}}\\I_{sum}^{-}\sin{\phi_{sum}^{-}}\\I_{sum}^{+}\cos{\phi_{sum}^{+}}
\end{bmatrix}}_\text{$I^{AC}$} -2U_{diff}^{0DC}
\underbrace{\begin{bmatrix}
I_{sum}^{aDC}\\I_{sum}^{bDC}\\I_{sum}^{cDC}
\end{bmatrix}}_\text{$I^{DC}$}
\label{eq:matrix}
\vspace{-0.3cm}
\end{equation}

\noindent where,

\vspace{-0.3cm}
{\small
\begin{align*}\nonumber
&M_{11} = Z_{arm}\left(-I_{s}^{+}\cos{\left(\rho-\phi_{s}^{+}\right)}-I_{s}^{-}\cos{\left(\rho-\phi_{s}^{-}\right)}\right)\nonumber-\\ &\nonumber -2U_{diff}^+\cos{\left(\theta_{diff}^{+}\right)}-2U_{diff}^-\cos{\left(\theta_{diff}^{-}\right)}
\end{align*}
\vspace{-0.5cm}
\begin{align*}\nonumber
&M_{12} = Z_{arm}\left(I_{s}^{+}\sin{\left(\rho -\phi_{s}^{+}\right)}+I_{s}^{-}\sin{\left(\rho-\phi_{s}^{-}\right)}\right)\nonumber-\\& \nonumber -2U_{diff}^+\sin{\left(\theta_{diff}^{+}\right)}-2U_{diff}^-\sin{\left(\theta_{diff}^{-}\right)}
\end{align*}
\vspace{-0.3cm}
\begin{align*}\nonumber
&M_{13} = M_{11}
\end{align*}
\vspace{-0.3cm}
\begin{align*}\nonumber
&M_{21} = Z_{arm}\left(I_{s}^{+}\cos{\left(\rho-\phi_{s}^{+} - \dfrac{2\pi}{3}\right)} -I_{s}^{-}\cos{\left(\rho-\phi_{s}^{-}\right)}\right)-\\& \nonumber-2U_{diff}^+\cos{\left(\theta_{diff}^{+}+\frac{2\pi}{3}\right)}-2U_{diff}^-\cos{\left(\theta_{diff}^{-}\right)}
\end{align*}
\vspace{-0.3cm}
\begin{align*}\nonumber
&M_{22} = Z_{arm}\left(-I_{s}^{+}\cos{\left(\rho-\phi_{s}^{+} - \dfrac{\pi}{6}\right)}+I_{s}^{-}\sin{\left(\rho-\phi_{s}^{-}\right)}\right)-\\& \nonumber-2U_{diff}^+\cos{\left(\theta_{diff}^{+}+\dfrac{\pi}{6}\right)}-2U_{diff}^-\sin{\left(\theta_{diff}^{-}\right)}
\end{align*}
\vspace{-0.3cm}
\begin{align*}\nonumber
&M_{23} = Z_{arm}\left(-I_{s}^{+}\sin{\left(\rho-\phi_{s}^{+}\right)}-I_{s}^{-}\cos\left(\rho-\phi_{s}^{-} + \frac{2\pi}{3}\right)\right)-\\& \nonumber - 2U_{diff}^-\cos{\left(\theta_{diff}^{-}-\frac{2\pi}{3}\right)}-2U_{diff}^+\cos{\left(\theta_{diff}^{+}\right)}
\end{align*}

\vspace{-0.3cm}
\begin{align*}\nonumber
&M_{31} = Z_{arm}\left(-I_{s}^{+}\cos{\left(\rho-\phi_{s}^{+} + \dfrac{2\pi}{3}\right)}-I_{s}^{-}\cos{\left(\rho-\phi_{s}^{-}\right)}\right)-\\& \nonumber-2U_{diff}^+\cos{\left(\theta_{diff}^{+}-\dfrac{2\pi}{3}\right)}-2U_{diff}^-\cos{\left(\theta_{diff}^{-}\right)}
\end{align*}
\vspace{-0.3cm}
\begin{align*}\nonumber
&M_{32} = Z_{arm}\left(I_{s}^{+}\cos{\left(\rho-\phi_{s}^{+}+\dfrac{\pi}{6}\right)}+I_{s}^{-}\sin{\left(\rho-\phi_{s}^{-}\right)}\right)+\\& \nonumber+2U_{diff}^+\cos{\left(\theta_{diff}^{+}-\dfrac{\pi}{6}\right)}-2U_{diff}^-\sin{\left(\theta_{diff}^{-}\right)}
\end{align*}
\vspace{-0.3cm}
\begin{align*}\nonumber
&M_{33} = Z_{arm}\left(-I_{s}^{+}\cos{\left(\rho-\phi_{s}^{+}\right)}-I_{s}^{-}\cos{\left(\rho-\phi_{s}^{-} - \dfrac{2\pi}{3}\right)}\right)-\\& \nonumber-2U_{diff}^-\cos{\left(\theta_{diff}^{-}+\dfrac{2\pi}{3}\right)}-2U_{diff}^+\cos{\left(\theta_{diff}^{+}\right)}
\end{align*}
}

Based on the vertical power references provided by the energy controllers, the AC additive current references can be obtained from \eqref{eq:matrix} as 
\vspace{-0.3cm}

{\small
\begin{align}
\label{eq:final_Ref}
&\begin{bmatrix}
I_{sum}^{-}\cos{\phi_{sum}^{-}}\\I_{sum}^{-}\sin{\phi_{sum}^{-}}\\I_{sum}^{+}\cos{\phi_{sum}^{+}}
\end{bmatrix} = 2U_{diff}^{0DC}
\begin{bmatrix}
I_{sum}^{aDC}\\I_{sum}^{bDC}\\I_{sum}^{cDC}
\end{bmatrix}\nonumber + \dfrac{1}{\det{M}} \\&
\footnotesize{\begin{bmatrix}
M_{22}M_{33}-M_{23}M_{32}&M_{13}M_{32}-M_{12}M_{33}&M_{12}M_{23}-M_{13}M_{22}\\
M_{23}M_{31}-M_{21}M_{33}&M_{11}M_{33}-M_{13}M_{31}&M_{11}M_{23}-M_{13}M_{21}\\
M_{21}M_{32}-M_{22}M_{31}&M_{12}M_{31}-M_{11}M_{32}&M_{11}M_{22}-M_{12}M_{21}\\
\end{bmatrix}}\nonumber\\&
\begin{bmatrix}
P_{u\to l}^{a}\\P_{u\to l}^{b}\\P_{u\to l}^{ac}
\end{bmatrix}
\end{align}}

The determinant of matrix $M$, during an internal singular voltage sag condition and considering that the AC grid current consists only of positive-sequence component, is equal to

\vspace{-0.3cm}

{\small
\begin{align}
\label{eq:sing}
& \det M  = -\dfrac{3Z_{arm}^3I_{s}^{+^3}\sqrt{3}\cos(\rho-\phi_{s}^{+})}{2} -
\\& \nonumber -3U_{diff}^+\sqrt{3}I_s^{+^2}Z_{arm}^2\cos(2\rho-2\phi_{s}^{+}+\theta_{diff}^{+})-\\& \nonumber-6I_s^+\sqrt{3}U_{diff}^{+^2}Z_{arm}\cos(2\theta_{diff}^{+}+\rho-\phi_{s}^{+})-\\& \nonumber {-6I_s^+\cos(\rho-\phi_{s}^{+})\sqrt{3}U_{diff}^{+^2}Z_{arm} -6I_s^{+^2}\cos(\theta_{diff}^{+})\sqrt{3}U_{diff}^+Z_{arm}^2} 
\end{align}
}

As some TSOs demand the injection of reactive currents to provide voltage support \cite{ELIA, 124} or active currents for frequency support \cite{NG} to the faulted phases throughout voltage sag events, the positive-sequence component of the AC grid current $I_s^+$ will generally be different than zero. Therefore, the suggested reference calculation will not present any discontinuities during internal singular voltage sag conditions, as it can be noted from \eqref{eq:sing}. Finally, the main characteristics of the different AC additive current reference calculation methods are summarized in Table \ref{tab:method_comp}.

\vspace{-0.3cm}
\begin{table}[ht]
\centering
\small
\caption{Methods summary}\renewcommand\arraystretch{1} 
\vspace{-0.3cm}
\begin{tabular}[c]{ccccccc}
\hline\hline
    \multirow{2}{*}{\textbf{Characteristics}} &
      \multicolumn{5}{c}{\textbf{Method}}\\
    & 0 & 1 & 2 &  3 & 4  \\
  \hline
    \multirow{1}{*}{MMC equivalent impedance}&\ding{53}  & \ding{53} & \ding{51} & \ding{51} & \ding{51} \\\hline
    \multirow{1}{*}{No energy drifts among phase-legs}& \ding{51}  & \ding{53} & \ding{51} & \ding{53} & \ding{51} \\\hline
    \multirow{1}{*}{Used for any voltage sag condition}& \ding{53}  & \ding{53} & \ding{53} & \ding{53} & \ding{51} \\\hline
    \multirow{1}{*}{Additive voltage effects}& \ding{53}  & \ding{53} & \ding{53} & \ding{53} & \ding{51} \\ \hline
    \multirow{1}{*}{Degrees of freedom are fully exploited}& \ding{53}  & \ding{53} & \ding{53} & \ding{53} & \ding{51} \\
\hline\hline
\end{tabular}
\label{tab:method_comp}
\vspace{-0.8cm}
\end{table}

\section{Case study} \label{sec:Results}

In this section, time-domain simulations are carried out in Matlab$\textsuperscript \textregistered$ Simulink to analyze the performance of the different reference calculation methods during AC grid (Section \ref{sec:AC_grid}) and internal singular (Section \ref{sec:int_grid}) voltage sag conditions. The simulations are performed considering an accelerated model of the MMC \cite{Xu} and employing the Nearest Level Control (NLC) technique to calculate the number of active sub-modules in each arm \cite{5673482}. In addition, the converter is considered to be operating under balanced AC grid conditions when the fault occurs. Both fault events last three seconds (starting at $t = 2$ s and restored at $t = 5$ s) \footnote{Note that for real networks, the maximum allowed time for fault-ride through would be equal to 250 ms \cite{Entso-e}.} in order to verify if the methods are able to keep the converter operational and to highlight the differences among them. Table \ref{tab:param_v2} details the system parameters for the case studies.
\vspace{-0.3cm}

\begin{table}[ht]
\centering
\small
\caption{System parameters}\renewcommand\arraystretch{1} 
\vspace{-0.3cm}
\begin{tabular}[c]{lccl}
\hline\hline
\textbf{Parameter}                   & \textbf{Symbol} & \textbf{Value} & \textbf{Units}     \\ \hline
Rated power                          & $S$               & 1000            & MVA                \\
Rated power factor                   & $\cos \phi$            & 0.95 (c)           & -           \\
AC-side rated voltage                      & $\underline{U}_{g}$&        325       & kV \\
HVDC link voltage                    & $U^{DC}$             & $\pm$320           & kV  \\
Phase reactor impedance              & $\underline{Z}_s$          & 0.005+j 0.18         & pu                 \\
Arm reactor impedance                & $\underline{Z}_{arm}$             & 0.01+j 0.15    & pu                 \\

Converter modules per arm            & $N_{u,l_{arm}}^k$            & 433            & -                  \\
Sub-module capacitance               & $C_{SM}$         & 9.5            & mF                \\

\hline\hline
\end{tabular}
\label{tab:param_v2}
\vspace{-0.5cm}
\end{table}

\vspace{-0.2cm}
\subsection{AC grid singular voltage condition} \label{sec:AC_grid}
\begin{figure}[!h]
\centerline{\includegraphics[width=1\linewidth]{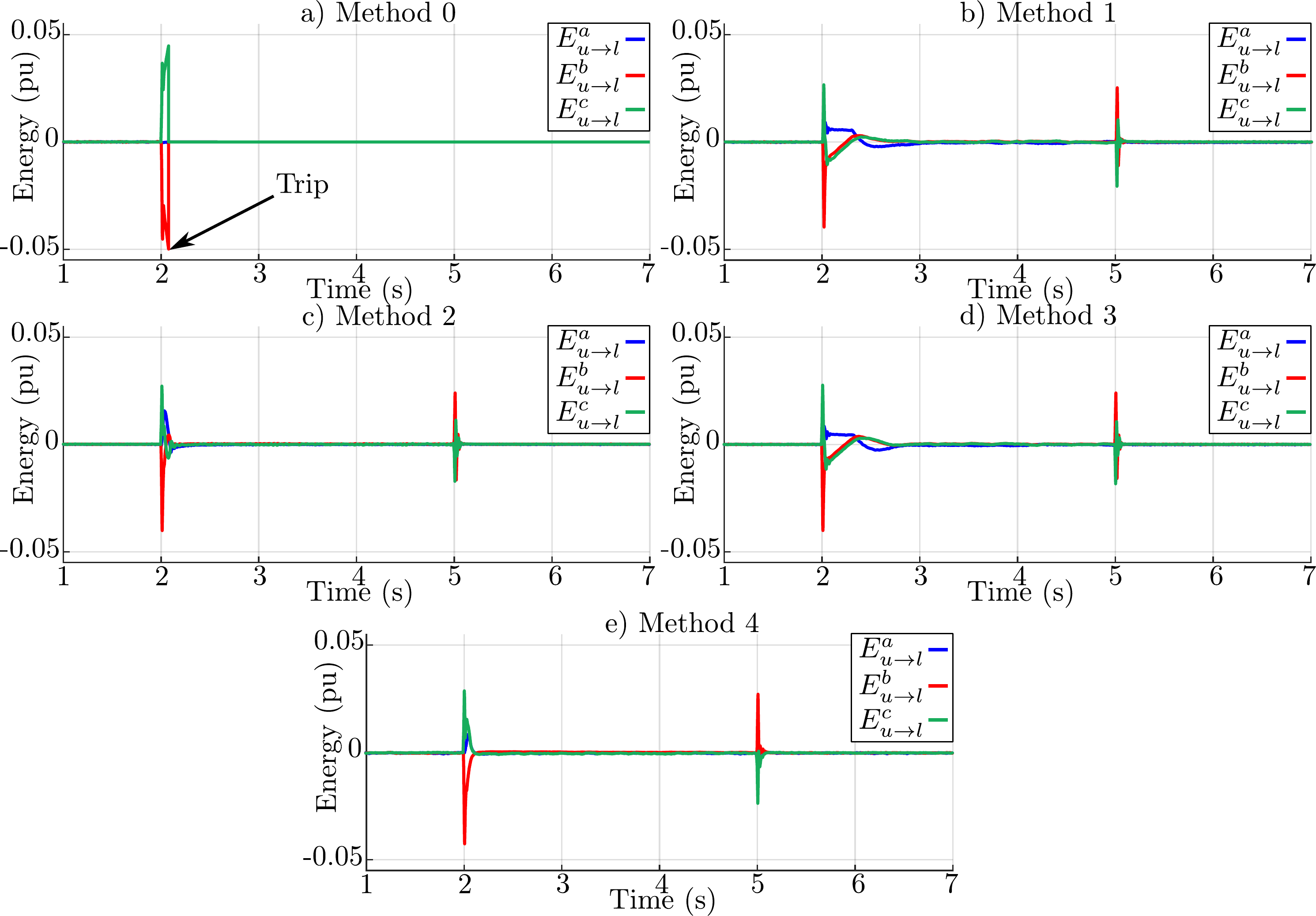}}
\vspace{-0.2cm}
\caption{Energy difference between the MMC arms during AC grid voltage singular condition.}
\label{fig:energy_ACsing}
\vspace{-0.5cm}
\end{figure}

\begin{figure}[!h]
\centerline{\includegraphics[width=0.75\linewidth]{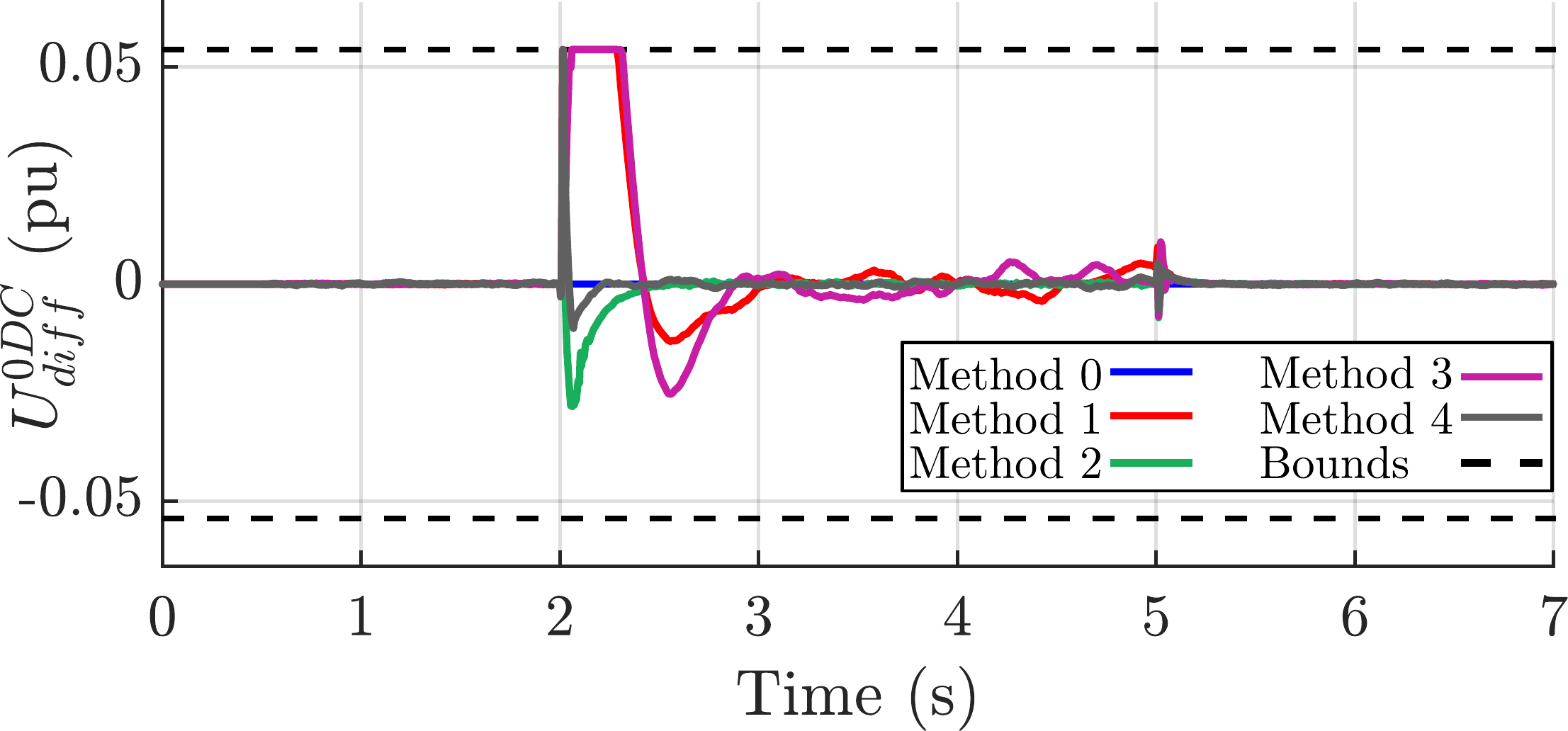}}
\vspace{-0.2cm}
\caption{$U_{diff}^{0DC}$ levels during AC grid singular voltage sag for the different reference calculation methods.}
\label{fig:Udiff_0DC_ACsing}
\vspace{-0.2cm}
\end{figure}

This case study is performed to illustrate the different dynamic behaviors that the converter will present according to the AC additive current reference calculation method used during an AC grid singular voltage condition type C \cite{edu_tran}. Under such fault event, the positive- and negative-sequence AC network voltage components have the same magnitude and phase-angle. In Fig. \ref{fig:energy_ACsing}, the MMC's internal energy transfer is shown for each phase throughout the simulation. It can be observed that only Method 0 leads to the disconnection of the converter which is in agreement with the theoretical analysis (see Section IV). Furthermore, when the fault occurs, a short sustained drift in the energy transfer between the arms of phase $a$ $(E_{u \to l}^a)$ for Methods 1 and 3 is noted. This happens because the DC differential zero-sequence voltage $U_{diff}^{0DC}$ controller saturates while attempting to eliminate the energy difference between the phase-legs of the converter, as shown in Fig. \ref{fig:Udiff_0DC_ACsing}. Otherwise, the modulation strategy might result in negative voltage levels to be applied into the SMs, which is not possible considering that half-bridge topologies are employed. If full-bridge SMs are considered, negative voltages could be imposed improving the dynamic response of the internal energy balancing.

\vspace{-0.3cm}

\begin{figure}[!h]
\centerline{\includegraphics[width=0.9\linewidth]{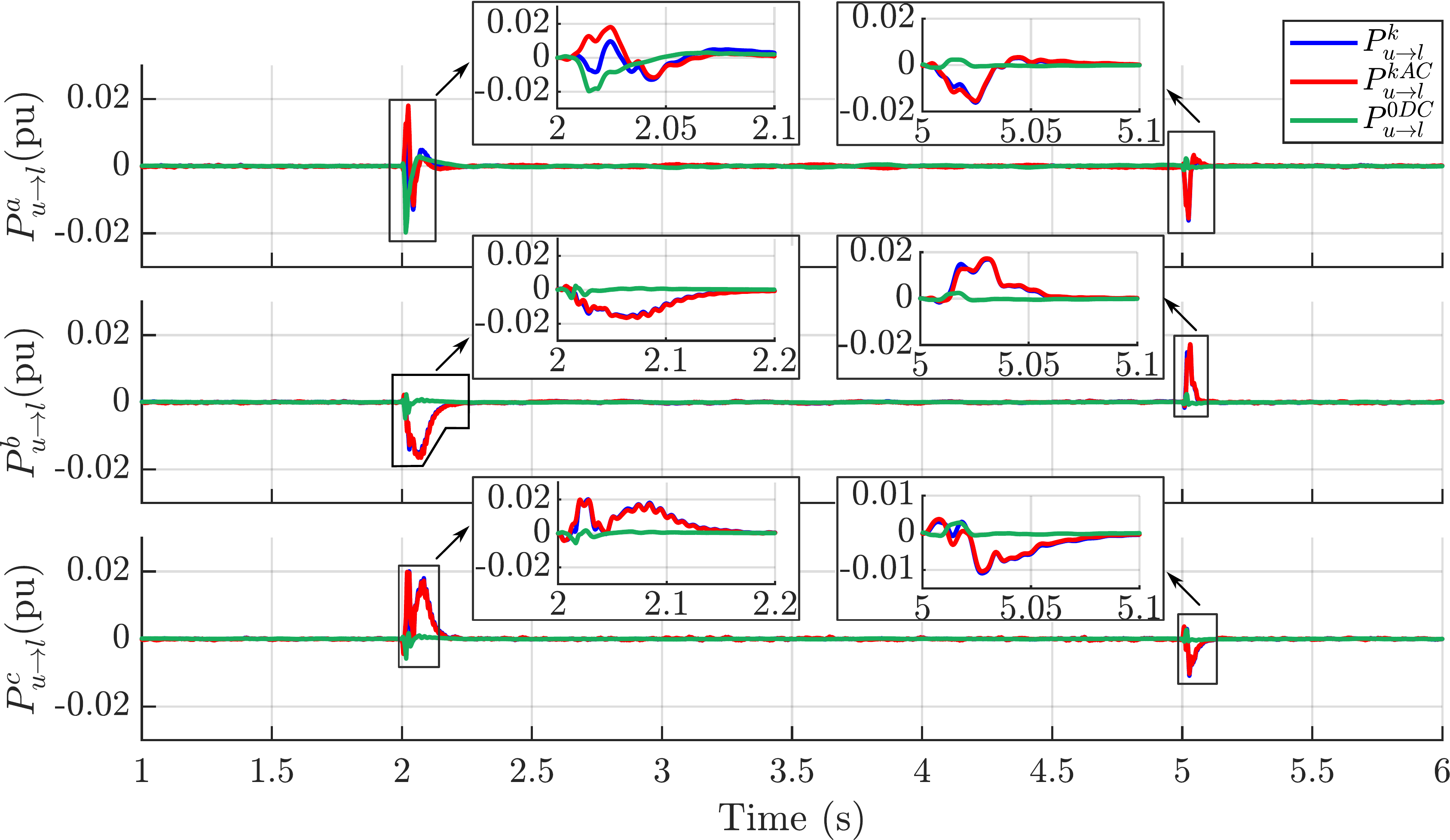}}
\vspace{-0.2cm}
\caption{Vertical power transfer during AC network singular voltage sag condition.}
\label{fig:power_balance}
\vspace{-0.5cm}
\end{figure}

In Fig. \ref{fig:power_balance}, the time-domain waveforms for the upper and lower power mismatches obtained employing \eqref{eq:Pu_l_abc_simp} are shown. The vertical power transfer regarding $U_{diff}^{0DC}$  is depicted in green, whereas the average AC power mismatches between the upper and lower arms is highlighted in red and the total value is represented as the blue continuous line. Among the fault occurrence and clearance transients highlighted in the figure, the DC component contribution is more evident during the fault event for phase $a$. Due to the characteristics of the fault and the AC grid current controller design, the active power injected in the faulted phases $b$ and $c$ is reduced, which is reflected inside the converter as a reduction in the DC additive circulating current levels for those phases. As a result, even though the DC zero-sequence voltage component is common for all the three phases, its effect is more significant for phase $a$ since its DC additive current level is maintained constant during the fault event. Finally, it can be better noted during the fault occurrence transient for phase $a$ that the power provided by $U_{diff}^{0DC}$ provides a negative component which reduces the high oscillations caused by the AC power part in the total vertical power transfer.

\subsection{Internal singular voltage condition} \label{sec:int_grid}
\begin{figure}[!h]
\centerline{\includegraphics[width=1\linewidth]{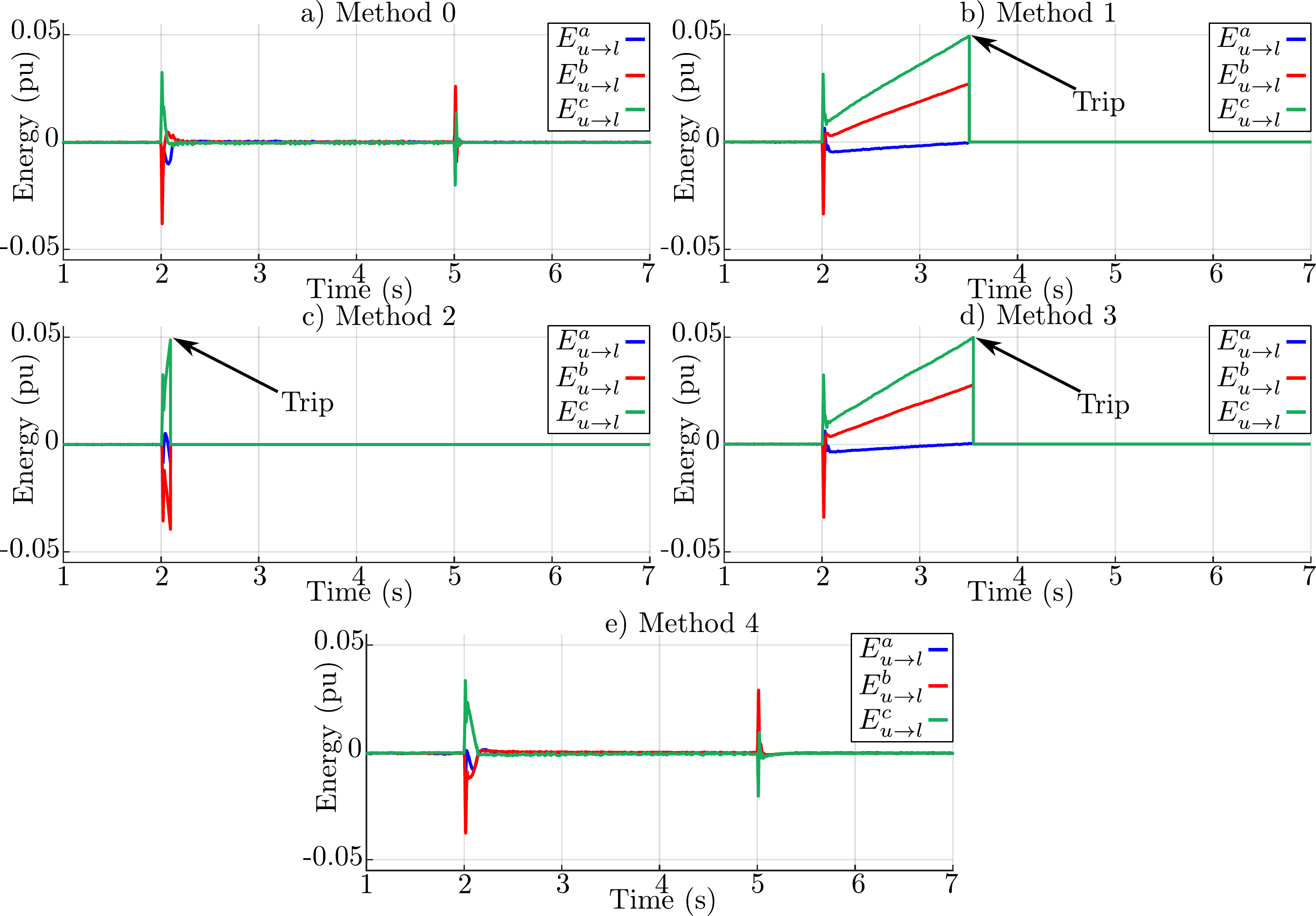}}
\vspace{-0.2cm}
\caption{Energy difference between the MMC arms during internal singular voltage sag condition.}
\label{fig:energy_internal}
\vspace{-0.2cm}
\end{figure}

The objective of this case study is to show that the proposed AC additive current reference calculation can avoid the discontinuity of the system even during internal singular voltage sags. The AC grid voltages for this fault are calculated based on the expression given in \eqref{eq:int_sing}, assuming an internal factor equals to $\underline{Z}_{eq}\underline{I}_s^{+}=0.24 \phase{87.75^o}$ pu and $\underline{U}_g^+ = 0.5\phase{0^o}$ pu, resulting in the positive- and negative-sequence AC differential voltages to be equal to $\underline{U}_{diff}^+=\underline{U}_{diff}^-=0.56\phase{25.49}^o$. The energy transfer profiles for each phase are shown in Fig. \ref{fig:energy_internal} for all the different reference calculation methods. It can be noted that, although Method 0 fails for AC grid singular voltage sag, it is capable of handling the internal singular voltage condition along with Method 4.

\begin{figure}[!h]
\centerline{\includegraphics[width=0.75\linewidth]{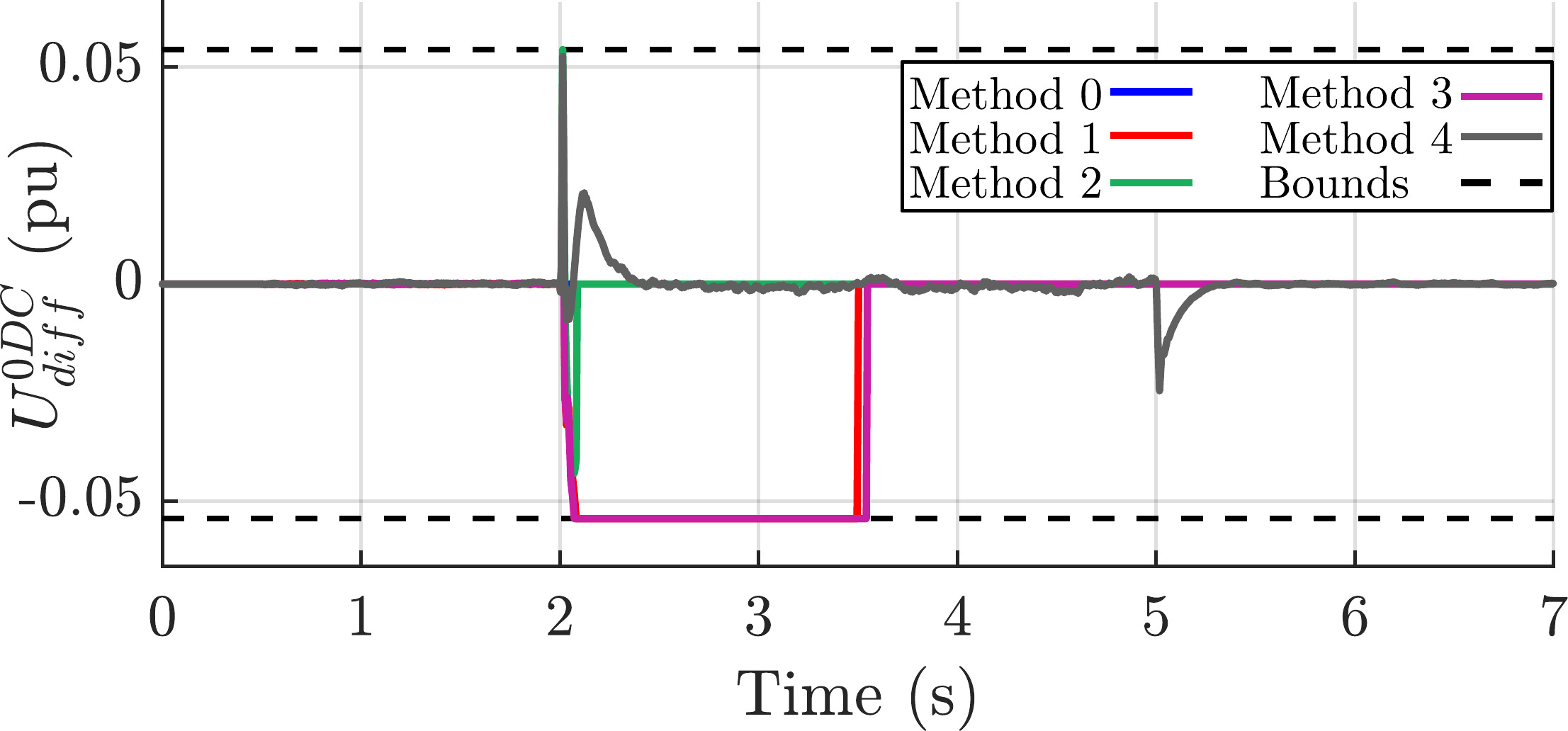}}
\vspace{-0.2cm}
\caption{$U_{diff}^{0DC}$ levels during internal singular voltage sag for the different methods.}
\vspace{-0.2cm}
\label{fig:Udiff_0_INTsing}
\vspace{-0.5cm}
\end{figure}

Regarding the DC zero-sequence differential voltage, Method 0 does not use it whereas all the other methods present either short or long saturation periods, as it can be seen in Fig. \ref{fig:Udiff_0_INTsing}. Method 2 is saturated for the maximum and minimum voltage levels, but it is not able to improve the energy regulation, leading to the system disconnection. Methods 1 and 3 result in a sustained saturation, but they also fail to track the desired energy references.
Although $U_{diff}^{0DC}$ is also saturated using Method 4, it is quickly recovered. 

\begin{figure}[!htb]
\centerline{\subfigure[Transition from normal operation to fault event.]{\includegraphics[width=1\linewidth]{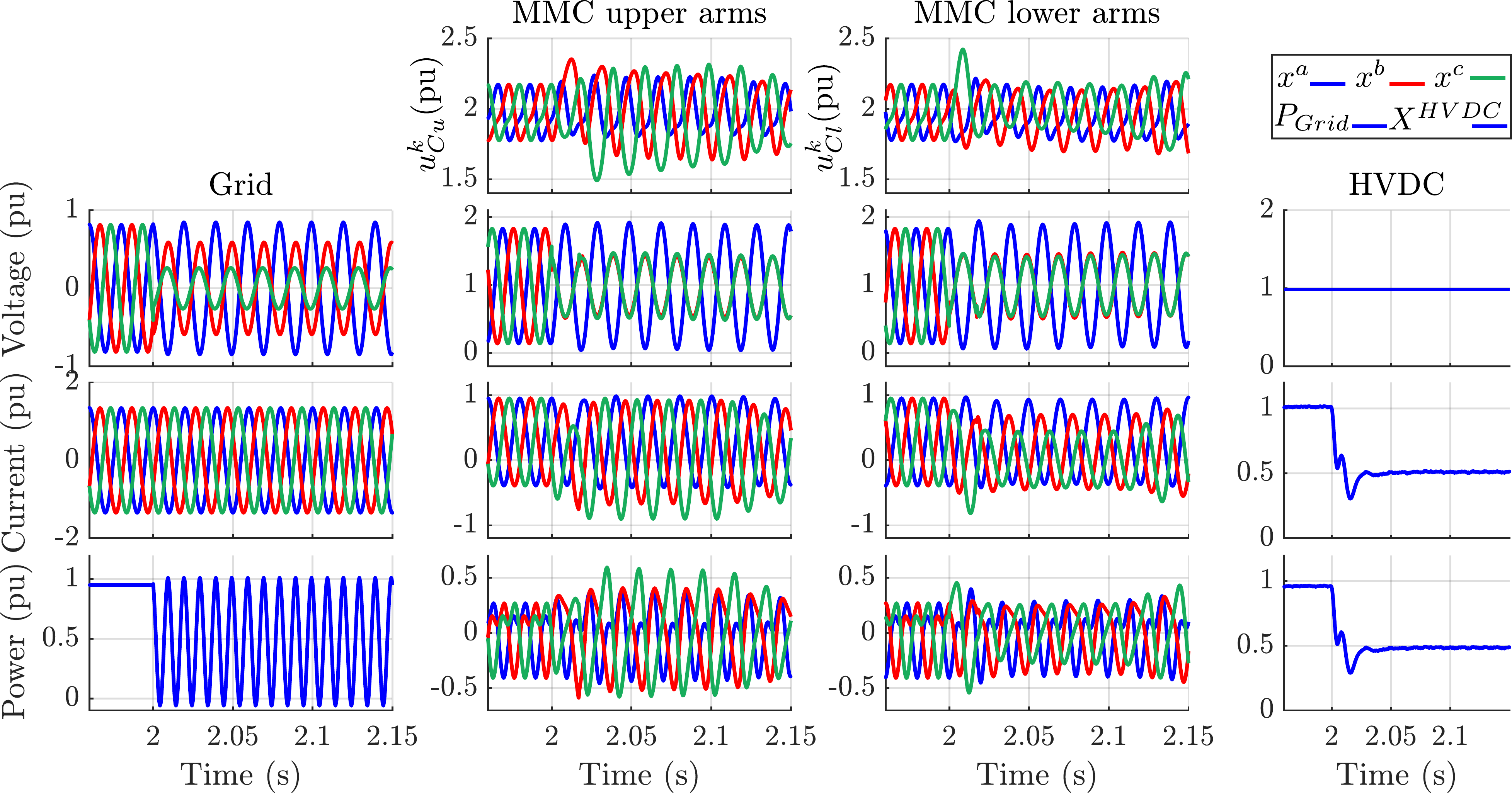}}} 

\centerline{\subfigure[Fault to normal.]{\includegraphics[width=1\linewidth]{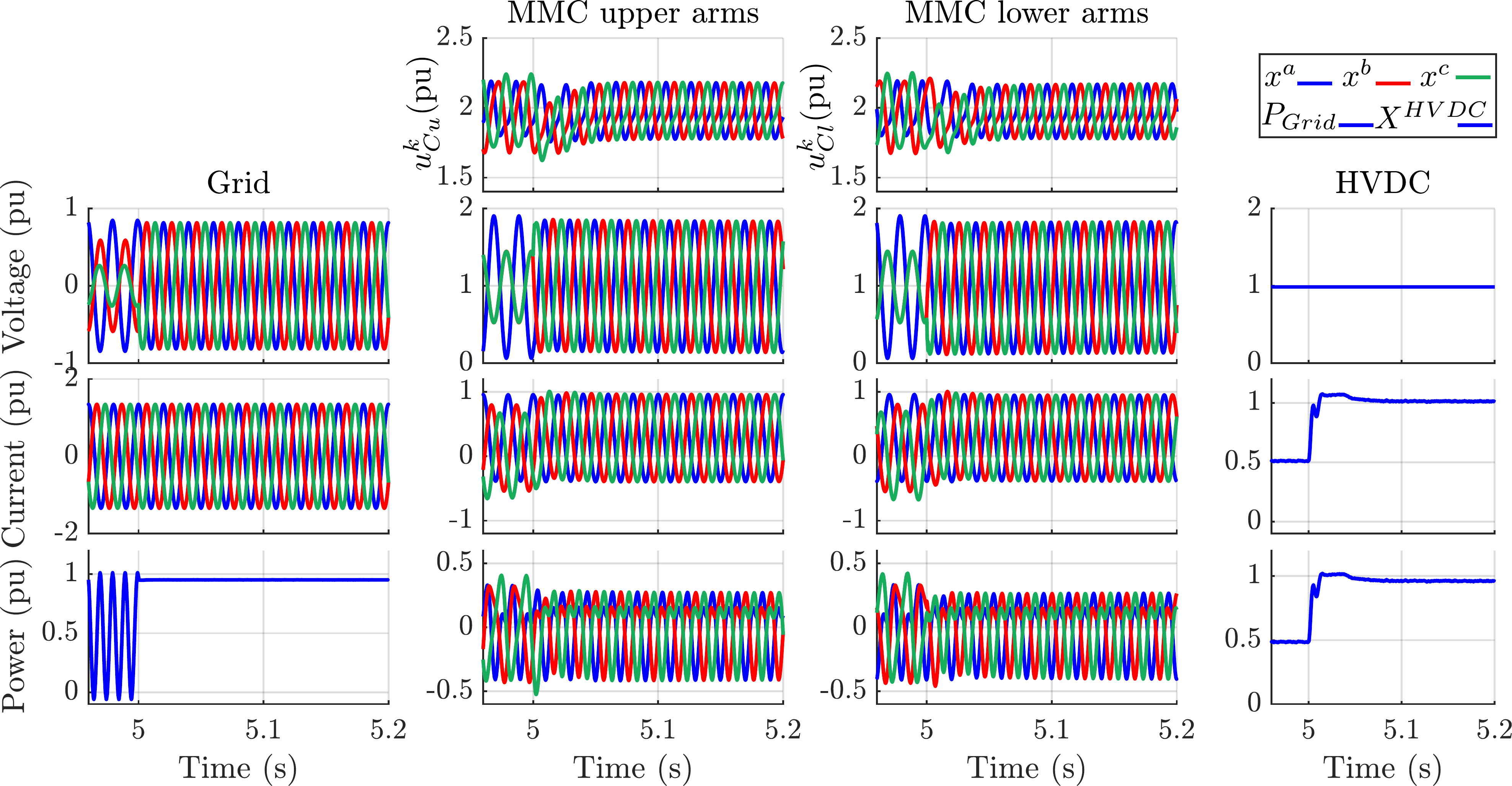}}}
\vspace{-0.2cm}
\caption{MMC waveforms during fault transients when Method 4 is employed. a) Fault is applied to the system and, b) Fault event is cleared.}
\label{fig:internal_waveforms}
\vspace{-0.5cm}
\end{figure}

In Fig. \ref{fig:internal_waveforms}, the waveforms of the MMC quantities are presented showing its dynamic behavior during the fault event and when it is cleared. It can be noted that the arms applied voltages for phase $b$ and $c$ become equal for this type of fault. In addition, all the voltages applied to the converter are higher than zero $(0 \leq U_{u,l}^k)$, which is obtained since the $U_{diff}^{0DC}$ is saturated. Finally, Method 4 was the only AC additive current reference calculation approach that avoided the converter to be tripped for both AC grid and internal singular voltage sag condition.

 \vspace{-0.2cm}

\subsection{Other singular fault scenarios}

\begin{figure}[!h]
\centerline{\includegraphics[width=1\linewidth]{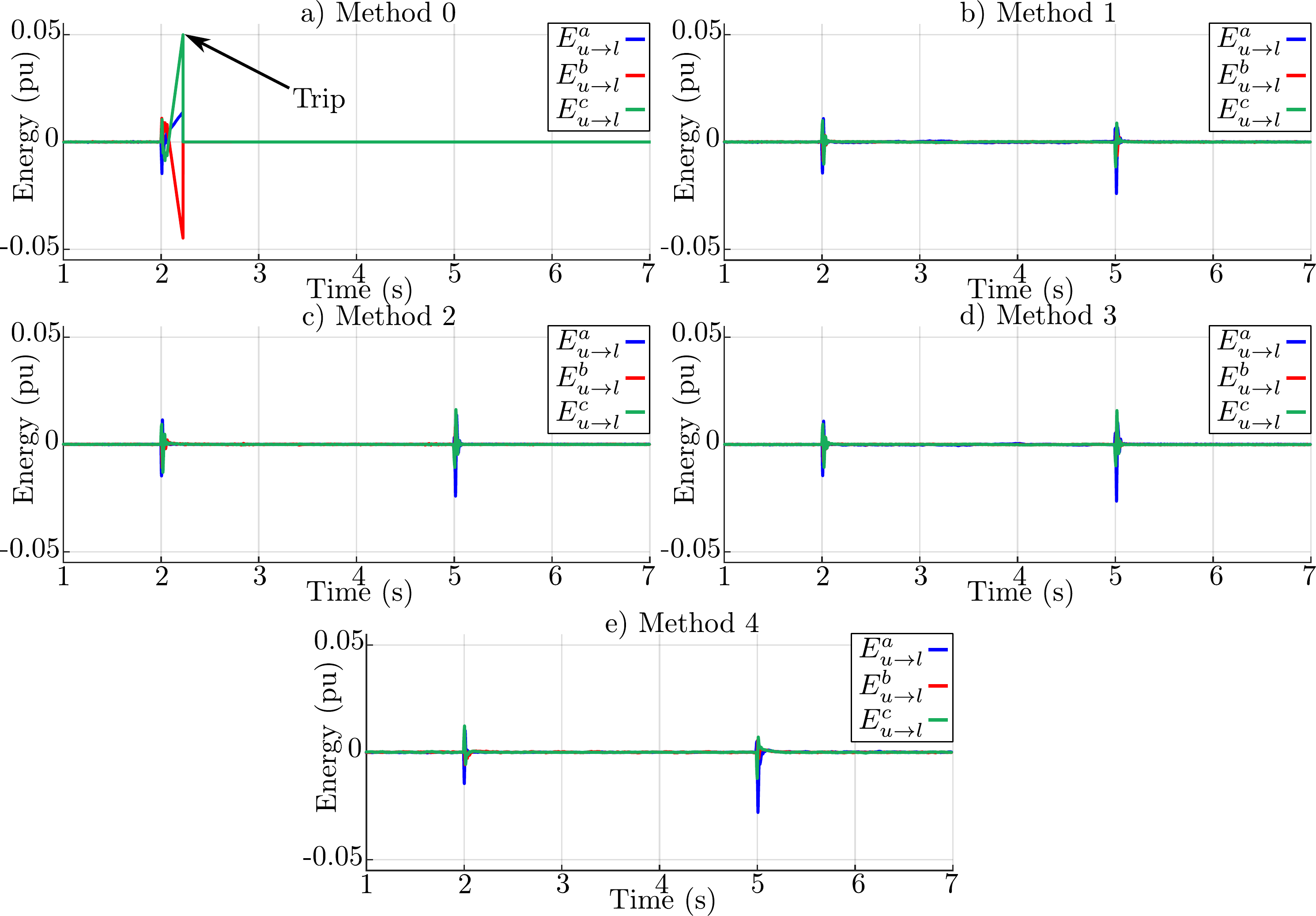}}
\vspace{-0.2cm}
\caption{Energy mismatches for AC network voltage singular type D.}
\label{fig:energy_AC_D}
\vspace{-0.6cm}
\end{figure}

\begin{figure}[!h]
\centerline{\includegraphics[width=1\linewidth]{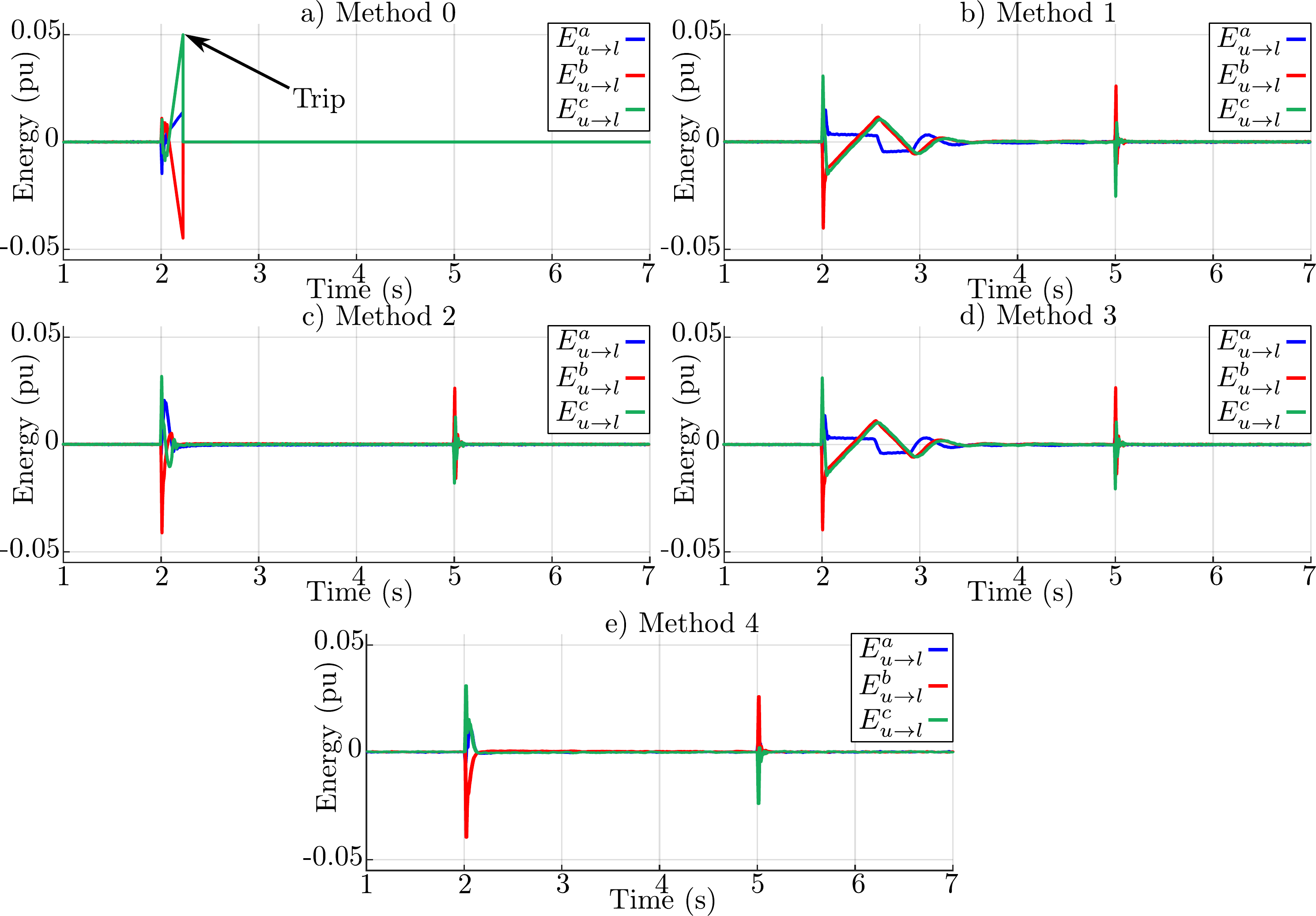}}
\vspace{-0.2cm}
\caption{Energy mismatches for AC network voltage singular type E.}
\label{fig:energy_AC_E}
\end{figure}

\begin{figure}[!h]
\centerline{\includegraphics[width=1\linewidth]{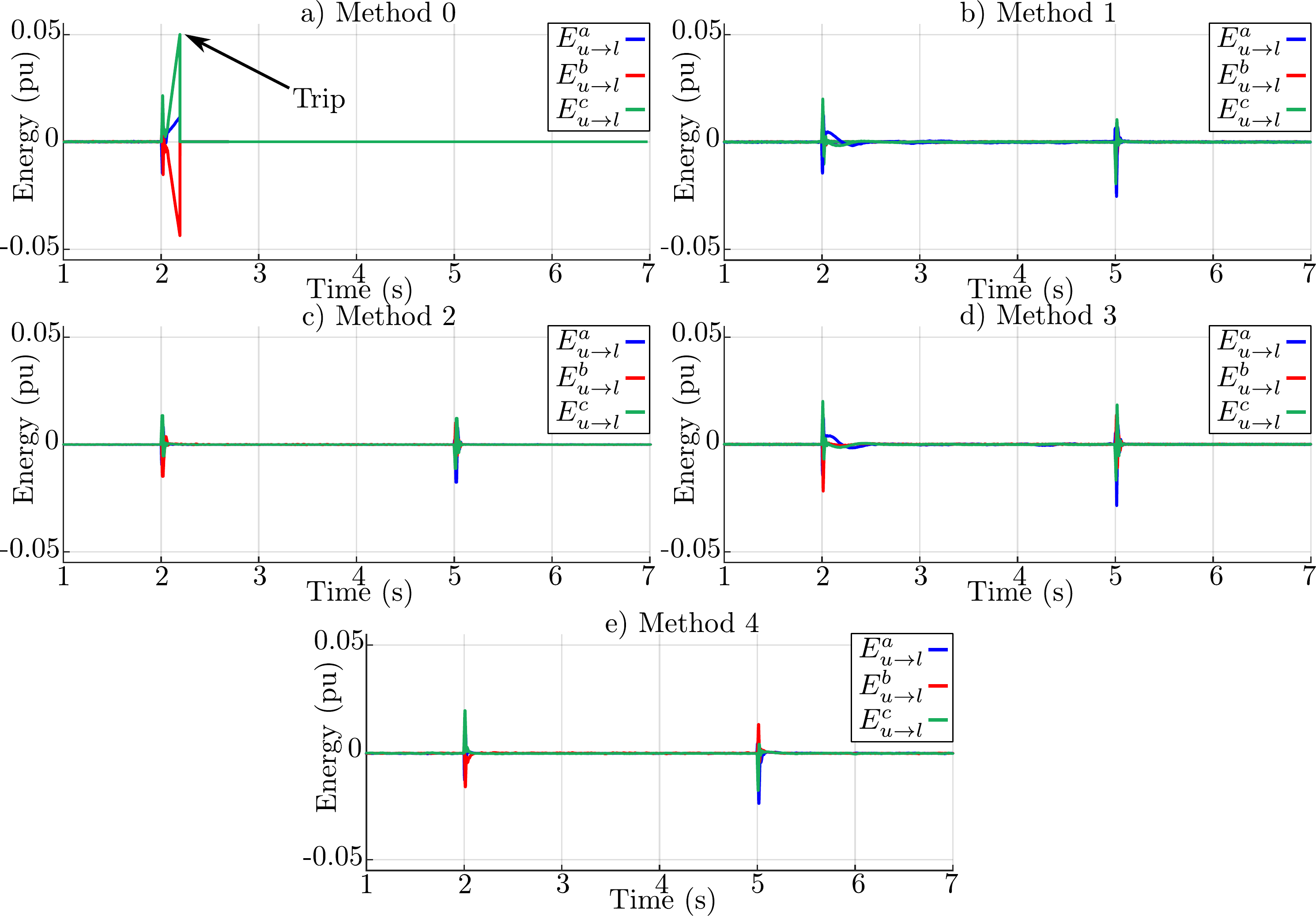}}
\vspace{-0.2cm}
\caption{Energy mismatches for AC network voltage singular type F.}
\label{fig:energy_AC_F}
\end{figure}

\begin{figure}[!h]
\centerline{\includegraphics[width=1\linewidth]{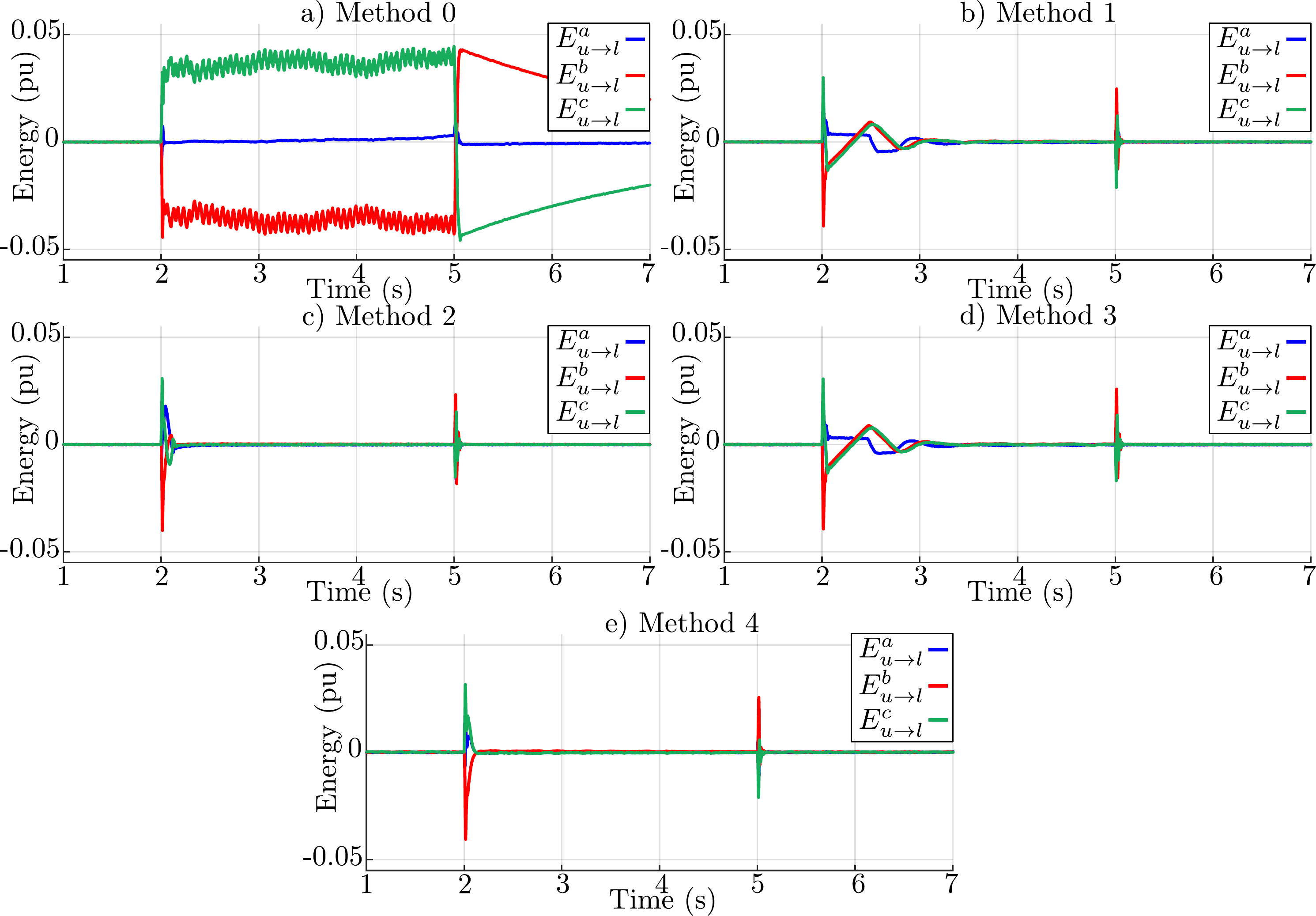}}
\vspace{-0.2cm}
\caption{Energy mismatches for AC network voltage singular type G.}
\label{fig:energy_AC_G}
\vspace{-0.5cm}
\end{figure}

\begin{figure}[!h]
\centerline{\includegraphics[width=1\linewidth]{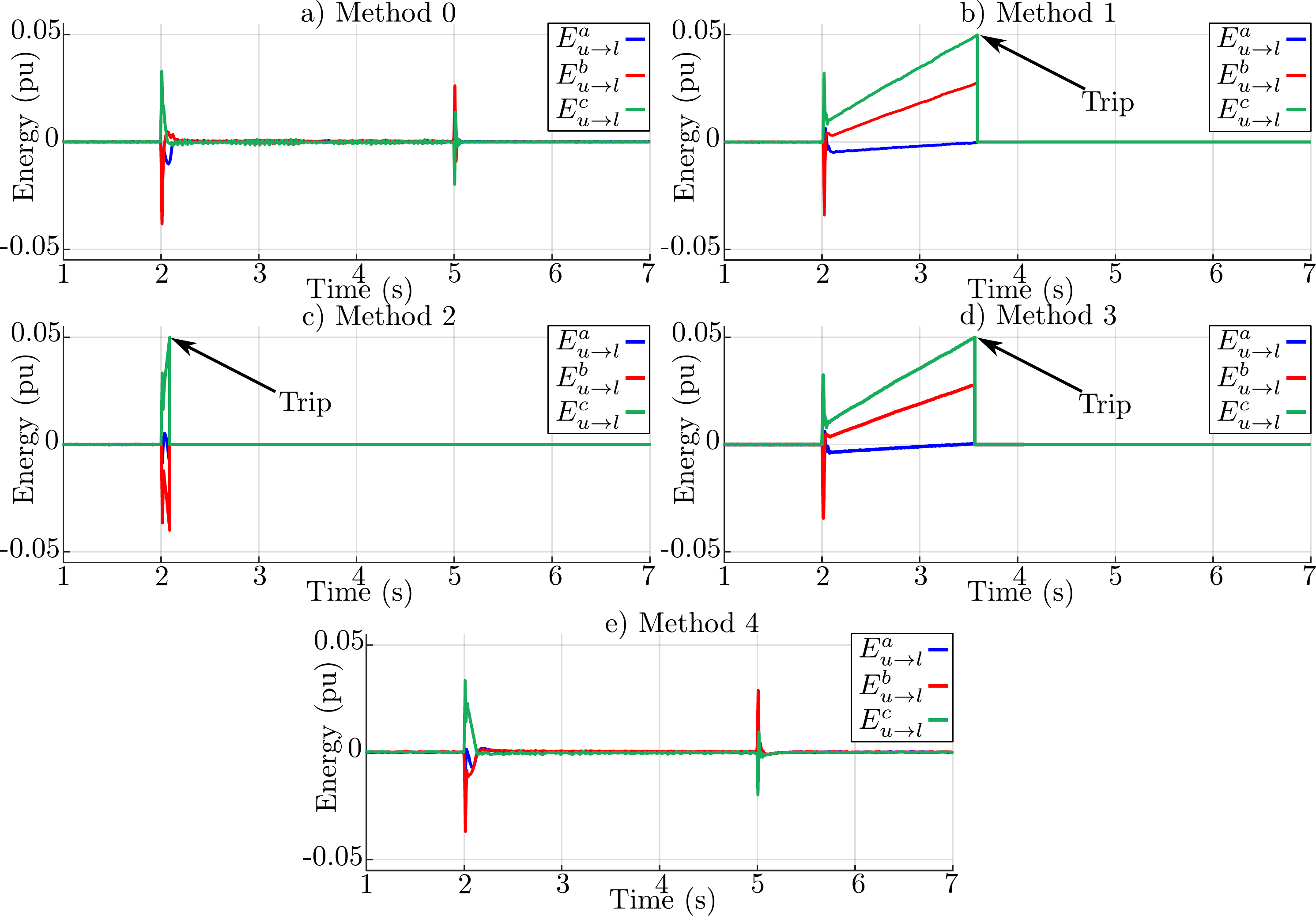}}
\vspace{-0.2cm}
\caption{Energy mismatches for internal voltage singular type  D.}
\label{fig:energy_internal_D}
\end{figure}

\begin{figure}[!h]
\centerline{\includegraphics[width=1\linewidth]{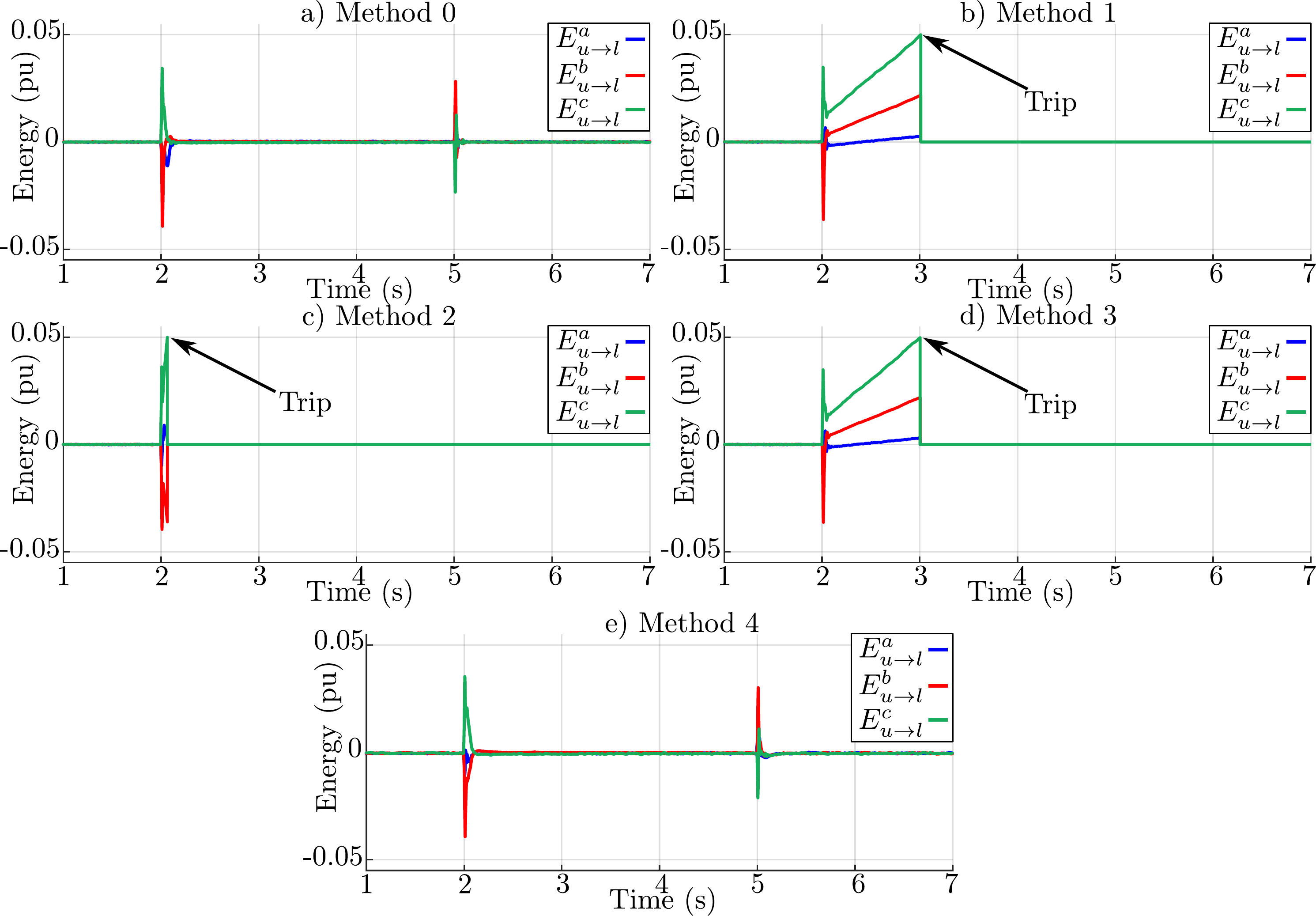}}
\vspace{-0.2cm}
\caption{Energy mismatches for internal voltage singular type  E.}
\label{fig:energy_internal_E}
\end{figure}

\begin{figure}[!h]
\centerline{\includegraphics[width=1\linewidth]{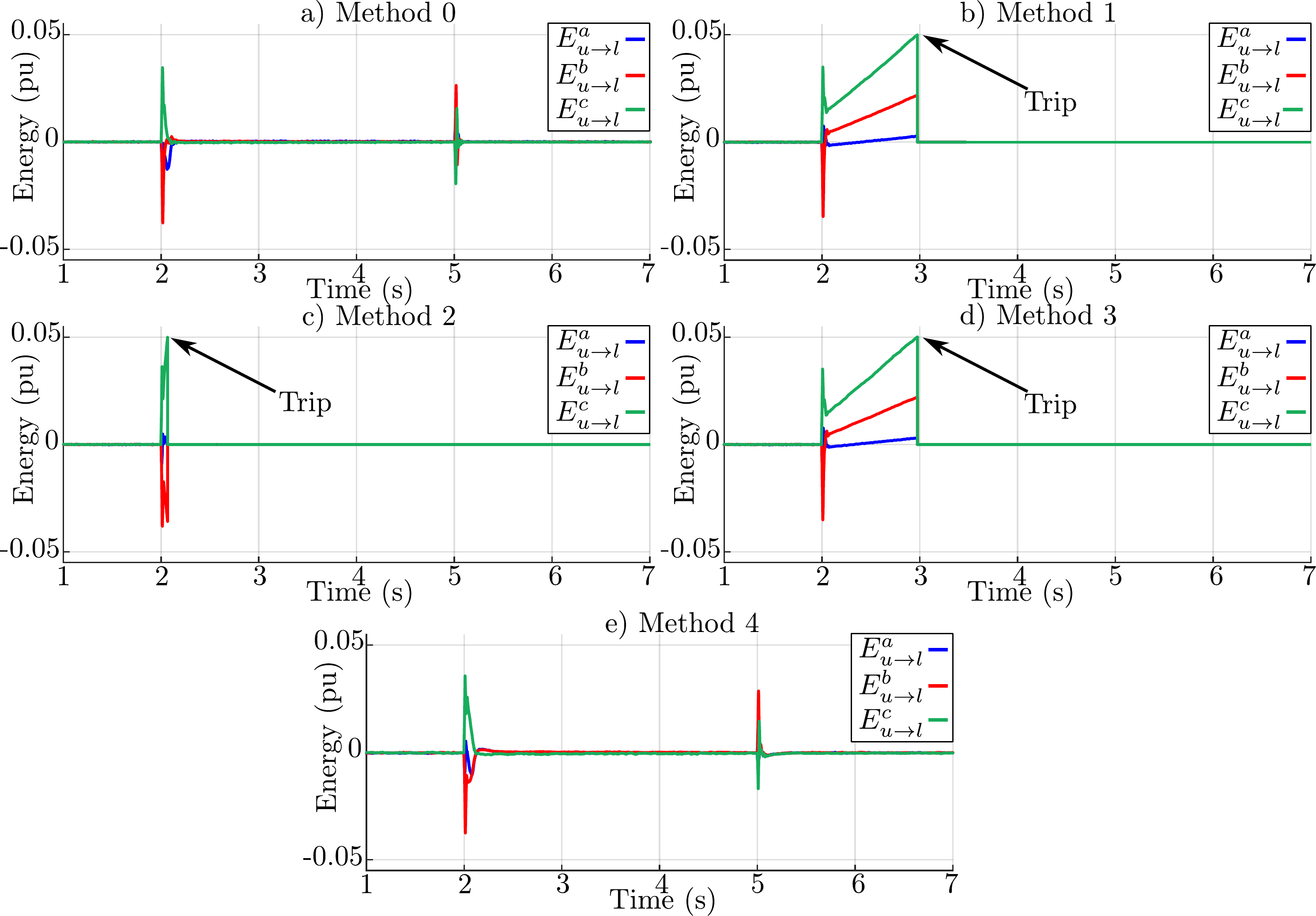}}
\vspace{-0.2cm}
\caption{Energy mismatches for internal voltage singular type  F.}
\label{fig:energy_internal_F}
\vspace{-0.5cm}
\end{figure}

\begin{figure}[!h]
\centerline{\includegraphics[width=1\linewidth]{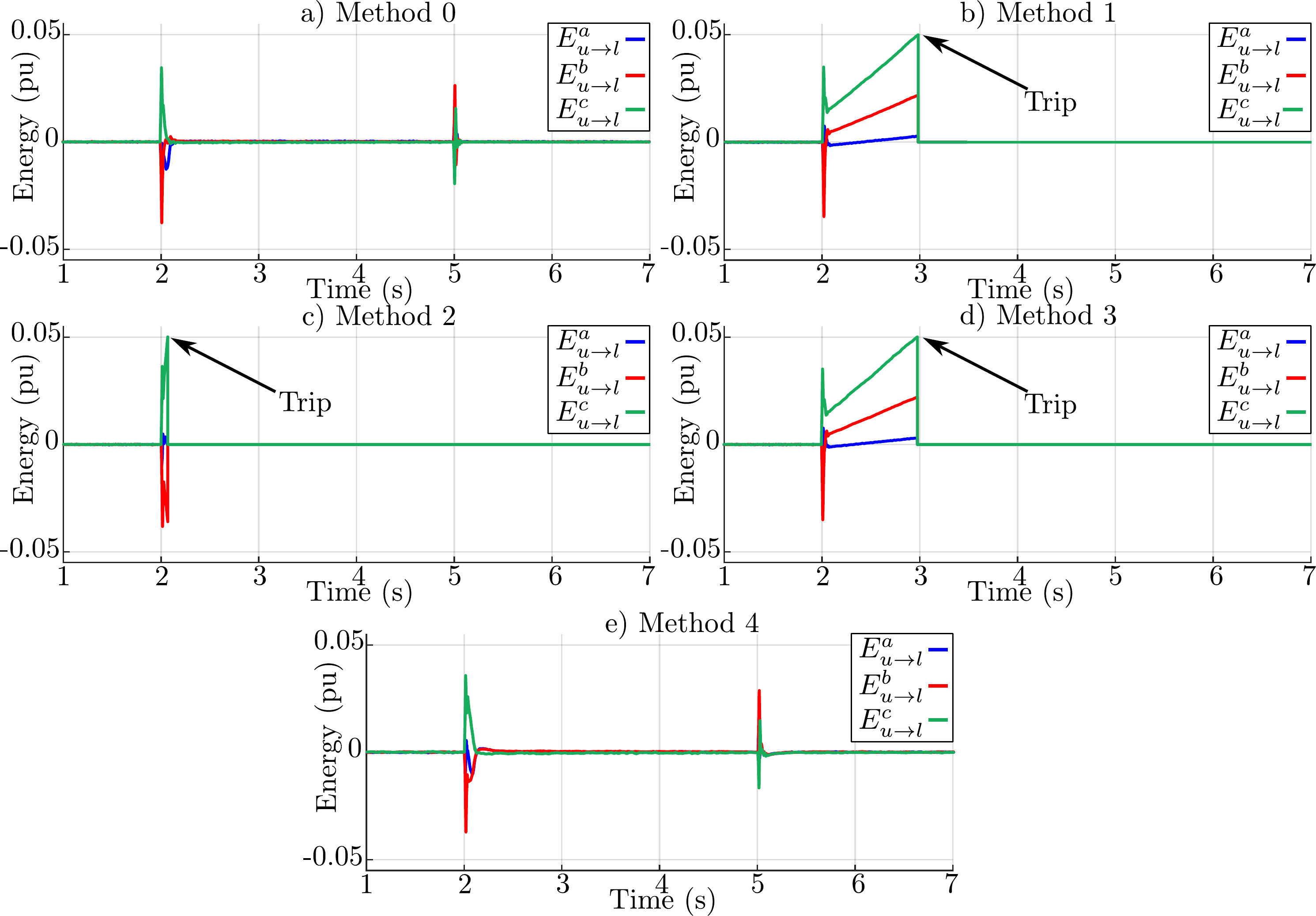}}
\vspace{-0.2cm}
\caption{Energy mismatches for internal voltage singular type  G.}
\label{fig:energy_internal_G}
\vspace{-0.5cm}
\end{figure}

In this section, the proposed method is compared with the other approaches for different types of internal and AC network singular voltage sag conditions. The simulations conducted focused on the upper and lower arms’ energy mismatch throughout the operation of the converter in order to further validate the proposed method. In Figs. \ref{fig:energy_AC_D} to \ref{fig:energy_AC_G} the results obtained during AC grid singular voltage sags D to G \cite{edu_tran} are shown, whereas Figs. \ref{fig:energy_internal_D} to \ref{fig:energy_internal_G} depict similar voltage sags, however reflected to the applied arm voltages of the converter characterizing $U_{diff}^+=U_{diff}^-$. 

The results confirm the conclusions drawn for the type C singular voltage sags. Regarding the AC grid singular conditions, it can be noted that Method 0 result in eventual disconnections of the converter (faults C to F), but it is able to marginally maintain the system operating during a type G fault. However, this method results in undesired sustained energy drifts during the aforementioned fault. In terms of Methods 1 and 3, during faults E and G, specifically, sustained energy drifts are observed for phase $a$, while the remaining phases present slow dynamics. Methods 2 and 4 have faster dynamics, quickly compensating the energy deviations.

During internal singular voltage conditions, Methods 1 to 3 are unable to regulate the converter, resulting in eventual protection trips (between 1 to 1.5 s after the fault’s occurrence for Methods 1 and 3 and within 100 ms for Method 2). On the other hand, Methods 0 and 4 are capable of managing the energy drifts caused by these faults. Finally, it should be mentioned that for all types of fault conditions presented, the proposed Method 4 was the only approach able to compensate the occurrences allowing the converter to safely reach steady-state conditions. 

\subsection{Internal parameters deviations}
In this case study, the performance of the proposed method 4 is analyzed considering parameters deviations in the arm impedances of the MMC during an internal singular voltage sag condition type D (see Section \ref{sec:internal} and \cite{edu_tran}). For the different internal parameter set-ups simulated, it is considered that both reference calculations and design of the controller gains are done based on the parameters given in Table \ref{tab:param_v2}. The effects of the arm impedance deviations are analyzed into two different set-ups. In the first one, unbalanced errors within $\pm5\%$ are considered, where in the second case the asymmetry can achieve errors up to $\pm10\%$. The arm impedance values for the different asymmetric cases are highlighted in Table \ref{tab:impedances}. Finally, the analysis is performed through time-domain simulations of the main quantities of the converter, as well as, its internal energy.

\begin{figure}[!htb]
\centerline{\subfigure[Transition from normal operation to fault event.]{\includegraphics[width=1\linewidth]{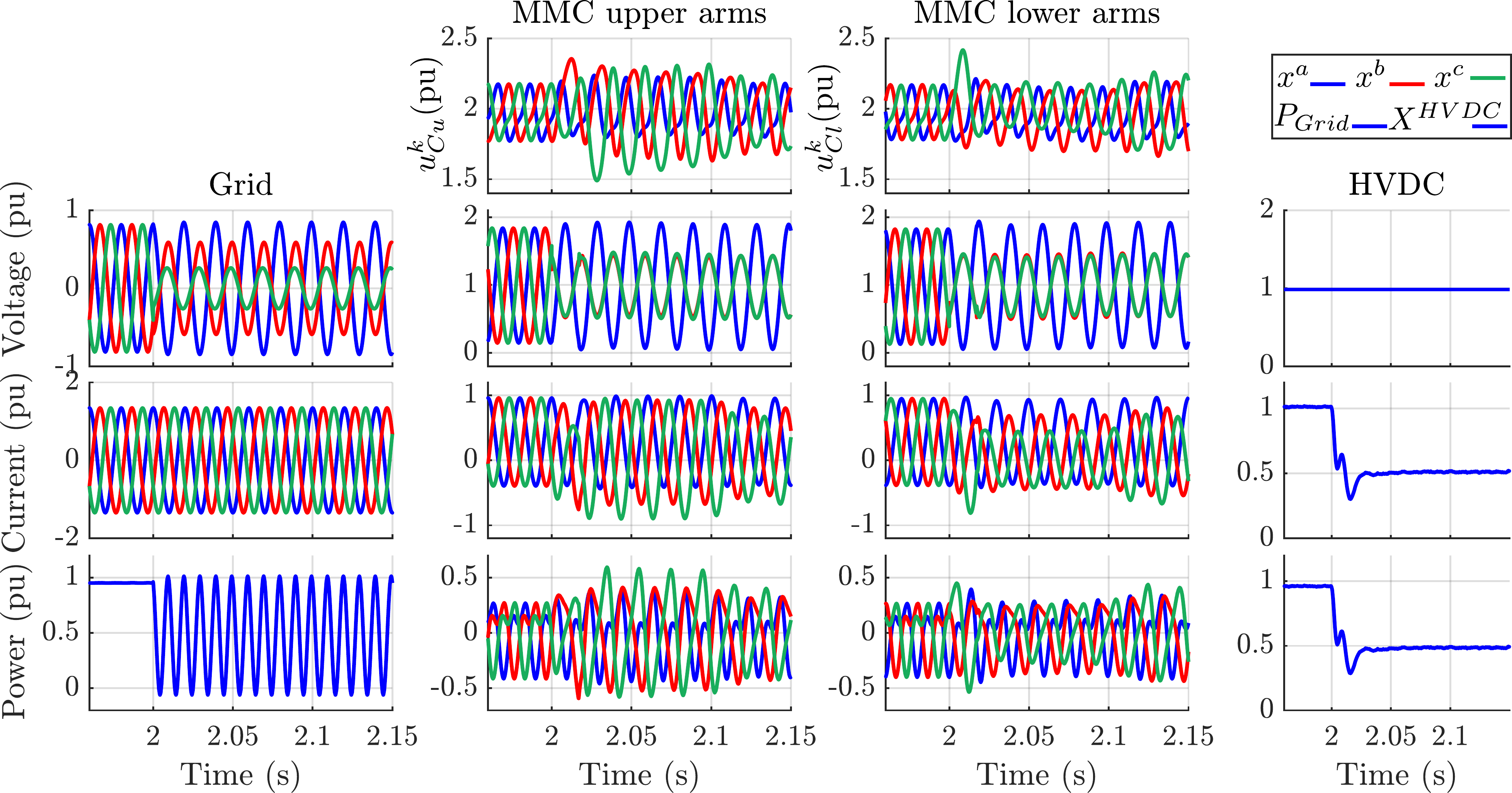}}} 

\centerline{\subfigure[Fault to normal.]{\includegraphics[width=1\linewidth]{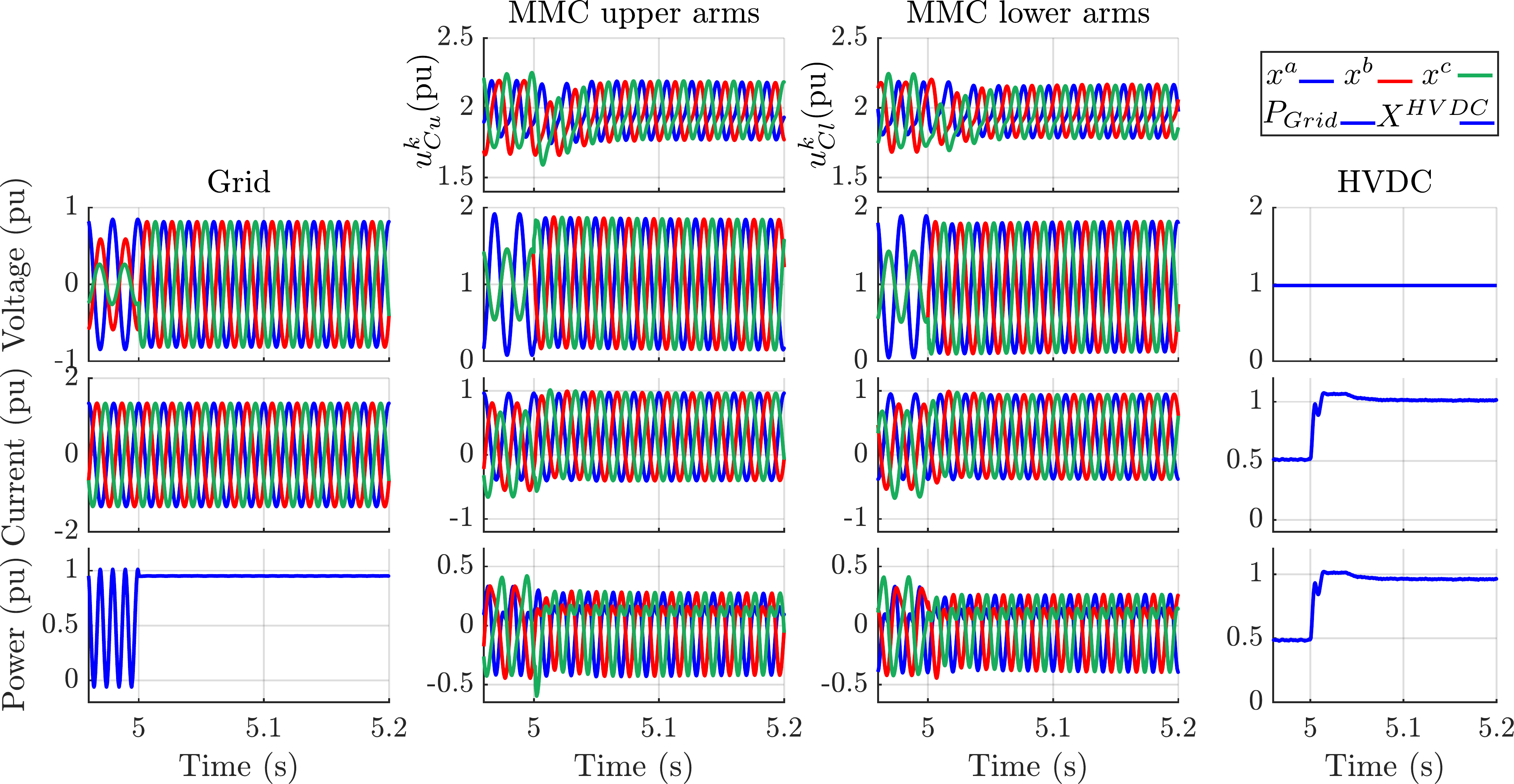}}}
\vspace{-0.2cm}
\caption{MMC waveforms during fault an interior singular voltage sag type D considering unbalanced arm impedances conditions within $\pm 5\%$ error . a) Fault is applied to the system and, b) Fault event is cleared.}
\label{fig:internal_waveforms_error5}
\vspace{-0.2cm}
\end{figure}

\begin{figure}[!h]
\centerline{\includegraphics[width=0.75\linewidth]{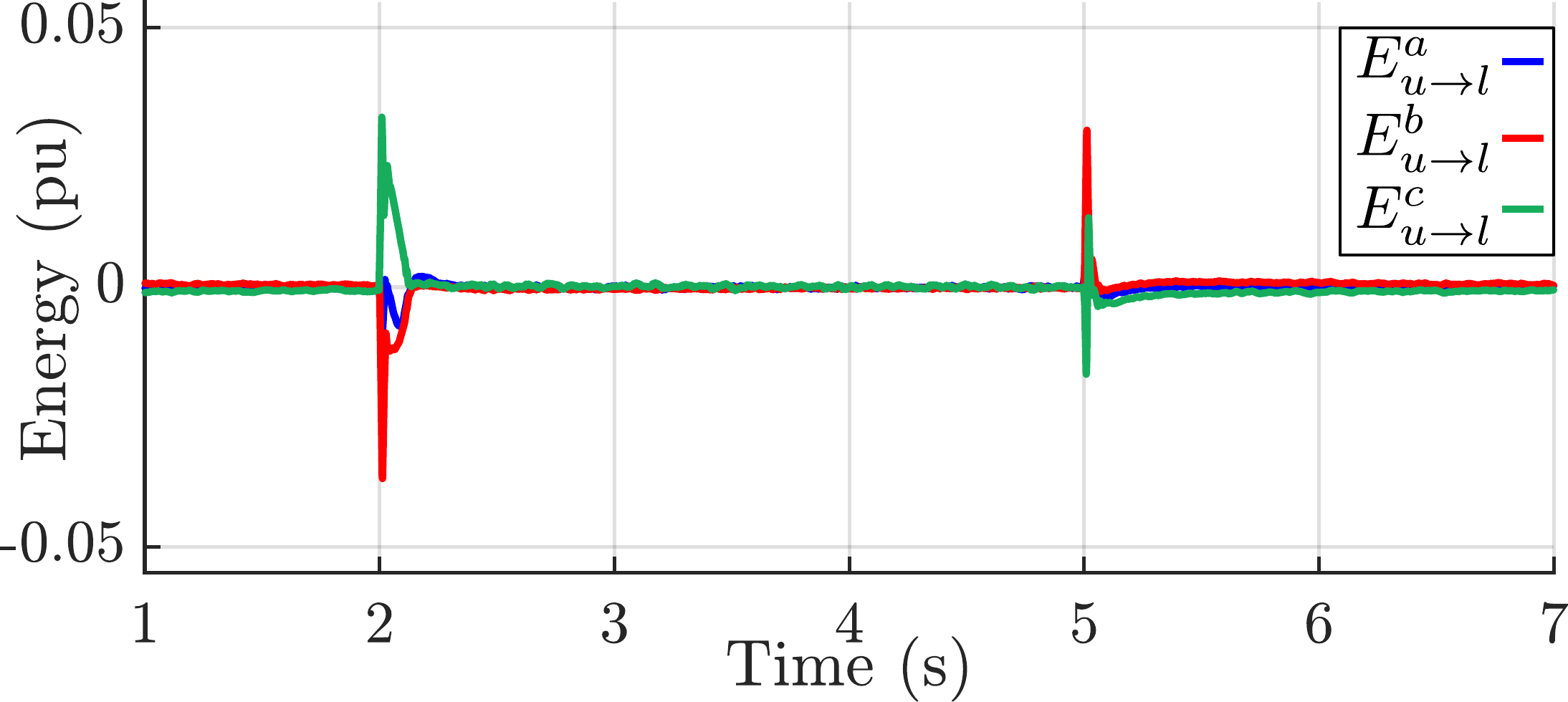}}
\vspace{-0.2cm}
\caption{Energy mismatches considering arm impedance unbalances within $\pm5\%$.}
\label{fig:energy_5e}
\vspace{-0.5cm}
\end{figure}

\begin{figure}[!htb]
\centerline{\subfigure[Transition from normal operation to fault event.]{\includegraphics[width=1\linewidth]{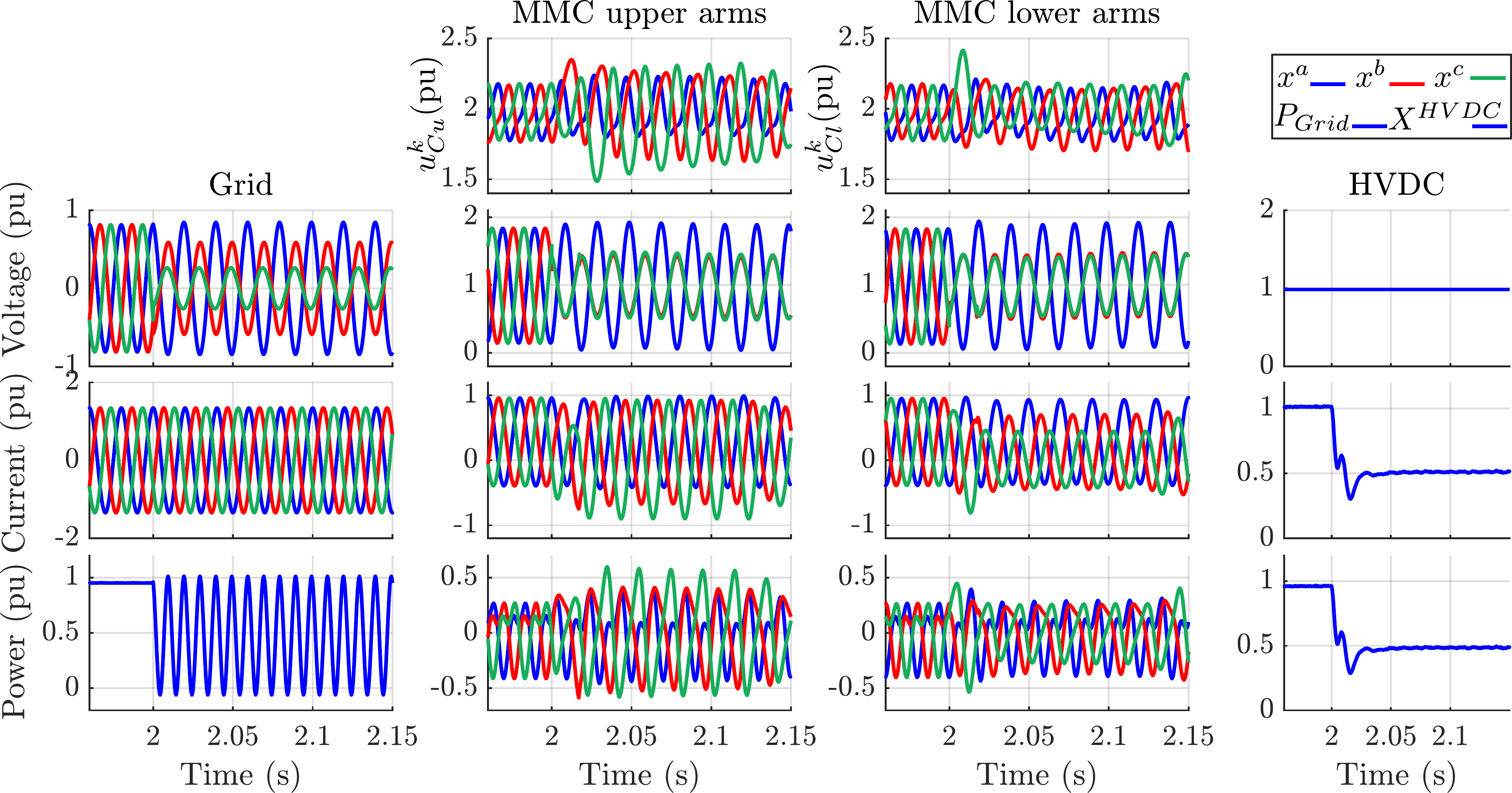}}} 

\centerline{\subfigure[Fault to normal.]{\includegraphics[width=1\linewidth]{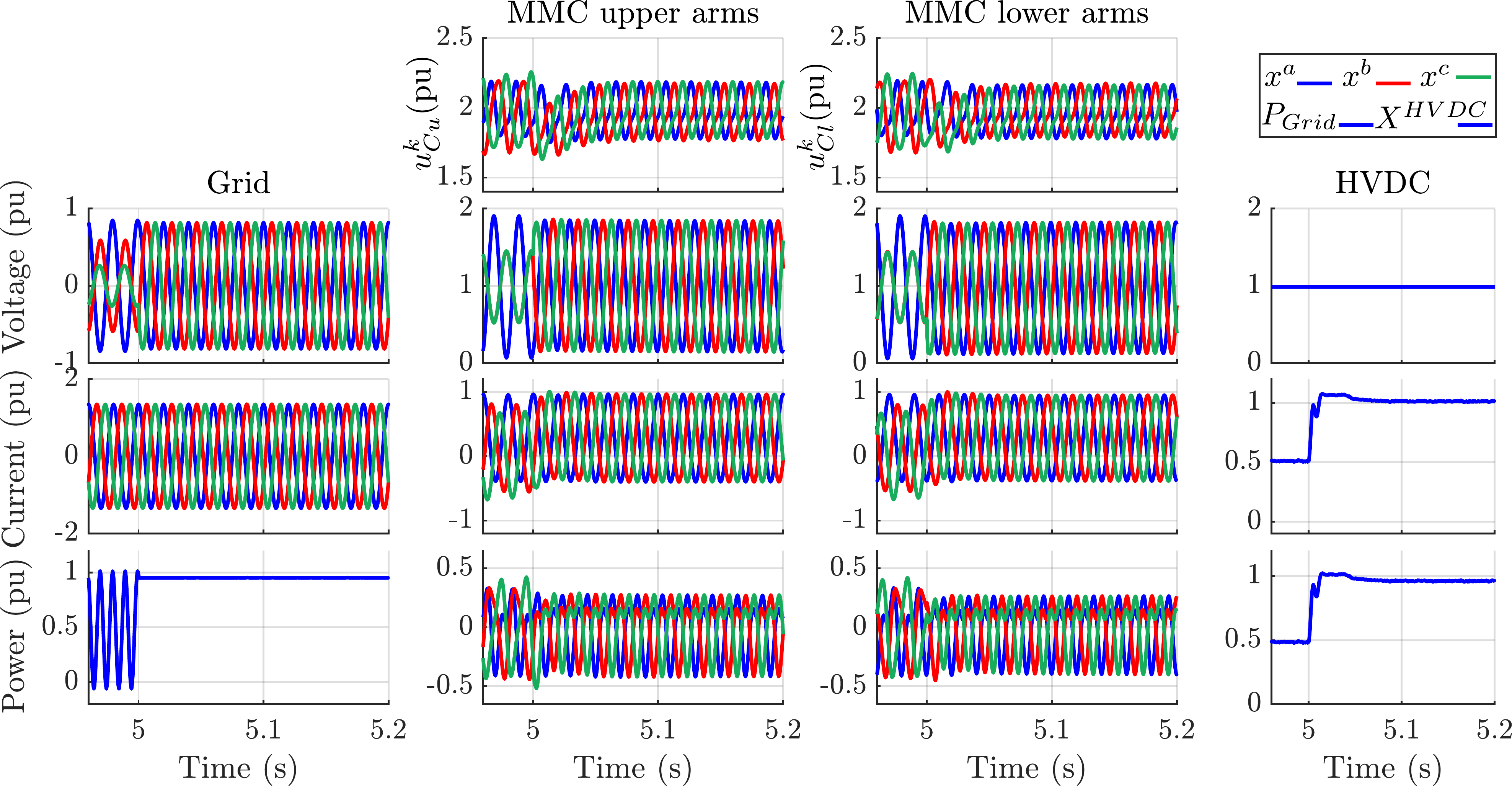}}}
\vspace{-0.2cm}
\caption{MMC waveforms during fault an interior singular voltage sag type D considering unbalanced arm impedances conditions within $\pm 10\%$ error . a) Fault is applied to the system and, b) Fault event is cleared.}
\label{fig:internal_waveforms_error10}
\vspace{-0.2cm}
\end{figure}

\begin{figure}[!h]
\centerline{\includegraphics[width=0.75\linewidth]{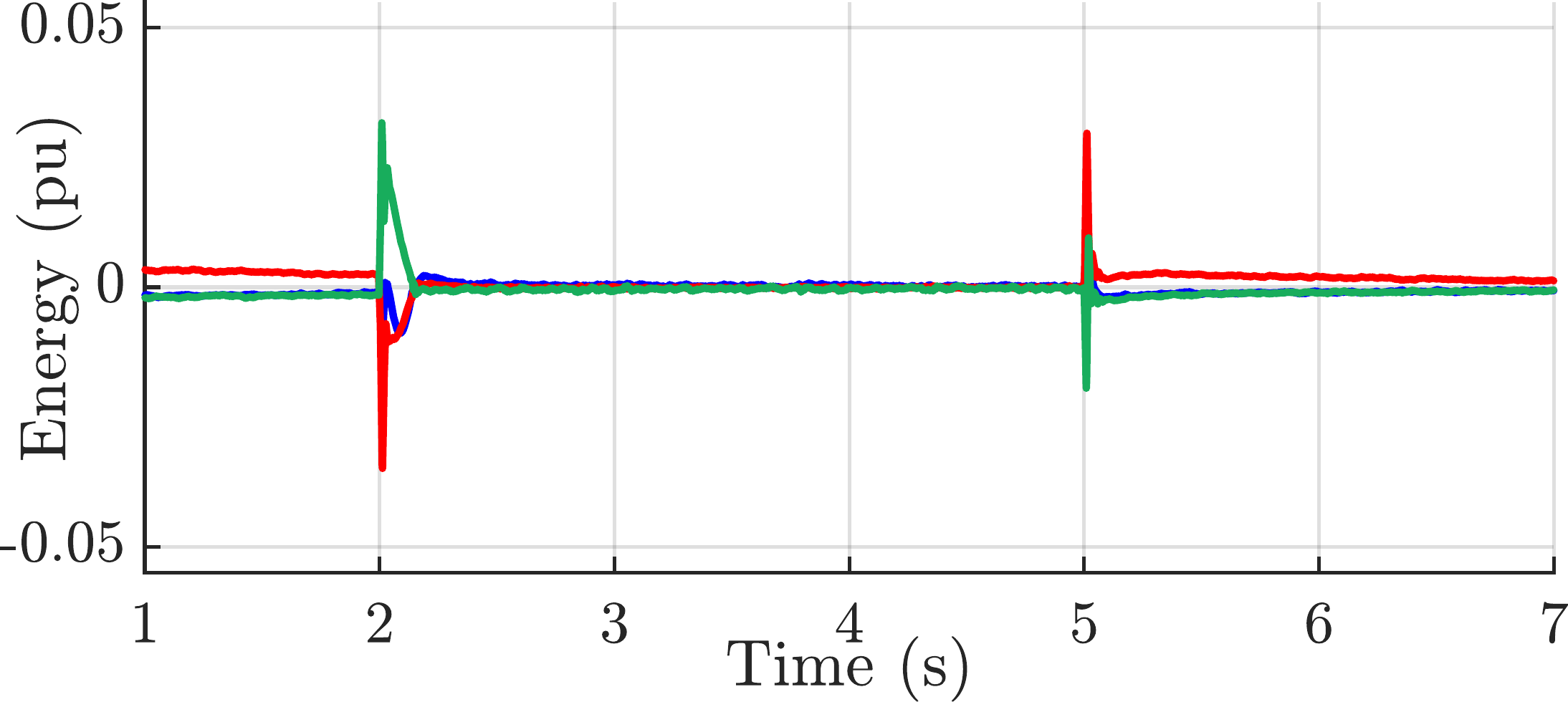}}
\vspace{-0.2cm}
\caption{Energy mismatches considering arm impedance unbalances within $\pm10\%$.}
\label{fig:energy_10e}
\vspace{-0.5cm}
\end{figure}

In Figs. \ref{fig:internal_waveforms_error5} and \ref{fig:energy_5e}, the waveforms for the $\pm5\%$ deviations are depicted. It can be noted that the arm impedance errors do not interfere with the proposed method, since it is still capable of maintaining the proper operation of the system even during the fault. Now, the errors are increased to $\pm10\%$ and the results are shown in Figs. \ref{fig:internal_waveforms_error10} and \ref{fig:energy_10e}. It can be observed that although the arm impedances present high asymmetric values, leading 100Hz oscillations in the power in the AC-side of the converter, such asymmetry does not affect the DC-side. If the reference calculations and the controllers are not properly designed, under such impedance condition, undesired 50Hz oscillations would appear in the DC-side current. Finally, comparing the energy plots during both unbalance scenarios, it is clear that the most severe case presents higher energy deviation among the phase-legs of the converter, but such deviation can be compensated during the steady-state and does not affect the overall performance of the MMC.

\vspace{-0.3cm}
\begin{table}[ht]
\centering
\small
\caption{Arm impedance values for Case study D}\renewcommand\arraystretch{1} 
\vspace{-0.3cm}
\begin{tabular}[c]{lccl}
\hline\hline
Deviation of $\pm5\%$                   &Deviation of $\pm10\%$     \\ \hline
$Z_{u}^{a} = Z_{arm} - 0.05Z_{arm}$   & $Z_{u}^{a} = Z_{arm} - 0.015Z_{arm}$                       \\
$Z_{u}^{b} = Z_{arm} - 0.01Z_{arm}$   & $Z_{u}^{b} = Z_{arm} - 0.1Z_{arm}$                  \\
$Z_{u}^{c} = Z_{arm} + 0.02Z_{arm}$                  & $Z_{u}^{c} = Z_{arm} + 0.13Z_{arm}$\\
$Z_{l}^{a} = Z_{arm} + 0.03Z_{arm}$            & $Z_{l}^{a} = Z_{arm} + 0.05Z_{arm}$              \\
$Z_{l}^{b} = Z_{arm} + 0.015Z_{arm}$           & $Z_{l}^{b} = Z_{arm} + 0.1Z_{arm}$                       \\
$Z_{l}^{c} = Z_{arm} + 0.025Z_{arm}$           & $Z_{l}^{c} = Z_{arm} - 0.08Z_{arm}$                       \\
\hline\hline
\end{tabular}
\label{tab:impedances}
\end{table}

\noindent where $Z_{u,l}^k$ is the upper and lower arms impedances, with $k \in \{a,b,c\}$.
 \vspace{-0.2cm}

 \vspace{-0.2cm}

\subsection{Distinctions among the reference calculation methods}
In this section, the main differences among the presented methods and the requirements for their implementation in a real system are discussed. The fundamental disparities among Methods 0 to 4 regard the consideration of the arms’ and AC network’s impedances effects and the usage of $U_{diff}^{0DC}$. Method 0 neglects both impedances and it is the only one that does not consider $U_{diff}^{0DC}$ in its vertical power equations. For Method 1, complex mathematical techniques to remove the degrees of freedom that do not contribute in the power transfer are required but still it does not acknowledge the impedances effects. Methods 2 and 3 extended the techniques applied in Methods 0 and 1, respectively, by considering the impedance impacts in the differential voltages. In Method 4, the impedances contributions are respected not only for the differential voltages but also for the additive ones.

From an implementational perspective, the previous methods present contrasting degrees of complexity. Regarding hardware requirements (e.g. sensors for measurements), all methods share similar control structures and would require similar measurements. For the methods regulating the DC differential zero-sequence voltage, no extra sensors are needed, since such quantity is calculated based on existing measurements. Considering the computational aspect, the implementation of Method 0 is the easiest one among the presented approaches as the most complex mathematical operation required is the inversion of a 3x3 matrix. Method 2 presents a slightly higher complexity level compared to Method 0 due to two main factors; 1- the usage of the differential quantities and the DC additive currents (the magnitude and phase-angle of the differential voltages, as well as, the DC current magnitudes can be obtained through basic operations and digital filters performed internally by the micro-controller), 2- the regulation of $U_{diff}^{0DC}$. Methods 1 and 3, although use different voltages in the calculations (AC grid and the AC differential voltages, respectively), they both require to compute the Moore-Penrose pseudoinverse in order to obtain their current references. Such complex mathematical operation is not required by Method 4. In this Method, the same procedures employed in Method 2 to obtain the differential voltages and DC additive currents are used only requiring an extra operation in the micro-processor to obtain the magnitudes and phase-angles of the AC additive currents. By having these values, the last operation required by the proposed Method 4 would be to solve equation \eqref{eq:final_Ref}.

In Table \ref{tab:Differences_methods}, the different measurements and the mathematical operations required by each method are highlighted. In summary, Method 0 is the most straight-forward and simplest to implement among all. Method 2 and 4 does not significantly increase the computational burden, whereas Methods 1 and 3 are the ones which require the highest computational burden, due to the Moore-Penrose pseudoinverse calculation. Finally, only Method 4 was capable of handling all the different fault case scenarios analyzed.

\begin{table}[!h]
\centering
\small
\caption{Distinction among Methods} \renewcommand\arraystretch{1} 
\begin{tabular}{cccccc}
\hline\hline
\multirow{2}{*}{\textbf{Additional operation and control}}                                                  & \multicolumn{5}{c}{\textbf{Method}} \\
                                                                                                   & 0   & 1   & 2   & 3   & 4  \\ \hline
Control of $U_{diff}^{0DC}$                                                                                   & \ding{53}   & \ding{51}    & \ding{51}    & \ding{51}    & \ding{51}   \\ \hline
\begin{tabular}[c]{@{}c@{}}Magnitude and phase-angle \\ of $u_{diff}^{+-}$\end{tabular}                  & \ding{53}   & \ding{53}   & \ding{51}    & \ding{51}    & \ding{51}   \\\hline
\begin{tabular}[c]{@{}c@{}}Calculation of the Moore-Penrose\\ pseudo matrix\end{tabular}         & \ding{53}   & \ding{51}    & \ding{53}  & \ding{51}    & \ding{53}  \\\hline
\begin{tabular}[c]{@{}c@{}}Extra matrix to remove\\ the AC additive current component\end{tabular} & \ding{53}   & \ding{51}    & \ding{53}   & \ding{51}    & \ding{53}  \\\hline
Arm impedance value                                                                                & \ding{53}  & \ding{53}   & \ding{53}   & \ding{53}   & \ding{51}   \\\hline
Magnitude and phase-angle of $u_g^{+-}$                                                                & \ding{51}    & \ding{51}    & \ding{53}   & \ding{53}   & \ding{53}  \\\hline
\begin{tabular}[c]{@{}c@{}}Inversion and solution \\ of 3x3 matrix\end{tabular}                   & \ding{51}   & \ding{51}    & \ding{51}    & \ding{51}    & \ding{51}  \\ \hline\hline
\end{tabular}
\label{tab:Differences_methods}
\vspace{-0.5cm}
\end{table}

\section{Conclusions}

An improved inner current reference calculation method for MMCs operating under normal and unbalanced network conditions has been presented. Particularly, the MMC operation during AC grid or internal singular voltage sag conditions. Such fault events are quite challenging to handle as they might lead to the eventual disconnection of the system by trying to impose high inner current references. The reference calculation has been formulated on the $abc$ additive and differential reference frames and it enables to calculate the arms' energy transfer considering all the degrees of freedom of the MMC; thus, the effects of the MMC's and AC side impedances are regarded. This is achieved through a mathematical substitution which does not require any iterative calculation method or optimization, whereby the AC additive voltage is replaced by the voltage drop in arms' impedance caused by the additive currents. Simulation results validate that the proposed reference calculation technique is able to provide the adequate converter references and to maintain the converter operating during both AC grid and internal singular voltage conditions.

\bibliographystyle{IEEEtran}

\bibliography{refs}

\end{document}